\documentclass{aa}  

\usepackage{graphicx}
\usepackage{txfonts}
\usepackage{lipsum}
\usepackage{subcaption}         
\usepackage{lscape}             
\usepackage{placeins}           

\usepackage{bm}
\usepackage{tabularx}
\usepackage{lscape}
\usepackage{stfloats}

\begin{document}

   \title{The $\sigma_s^2-\mathcal{M}$ relation in the multi-phase ISM} 

   \subtitle{Exploring the density PDF with the Cloud Factory simulations}


   \author{M. Nonhebel\inst{1 \and 2} \and R. J. Smith\inst{2} \and R. S. Klessen\inst{3}}

   \institute{Max Planck Institute for Astronomy, Königstuhl 17, 69117 Heidelberg, Germany \and SUPA, School of Physics and Astronomy, University of St. Andrews, North Haugh, St. Andrews KY16 9SS, UK \and Universität Heidelberg, Zentrum für Astronomie, Institut für Theoretische Astrophysik, Albert-Ueberle-Str 2, 69120 Heidelberg, Germany}
   \date{}

 
  \abstract
   {The density probability distribution (PDF) of molecular clouds is a crucial component of analytical theories of star formation. In idealised simulations of isothermal turbulence, the width of the density PDF,  $\sigma_s^2$, is dependent on the sonic Mach number of the medium, $\mathcal{M}$. The $\sigma_s^2-\mathcal{M}$ relation is widely used to connect cloud-scale turbulence to the density PDF, and further to star formation activity, yet its validity within individual phases of the multi-phase interstellar medium (ISM) remains untested.}
   {In this study, we evaluate whether the $\sigma_s^2-\mathcal{M}$ relation is applicable to individual phases of the ISM.} 
   {We study the density PDFs of molecular cloud complexes in the Cloud Factory simulations; a suite of detailed zoom-in simulations that self-consistently generate a turbulent, multi-phase ISM. We test whether the $\sigma_s^2-\mathcal{M}$ relation holds in the hot ionised medium (HIM), warm ionised medium (WIM), warm neutral medium (WNM), cold neutral medium (CNM), the molecular phase, and the highly-shielded molecular phase traced by CO.}
   {We find the applicability of the classical $\sigma_s^2-\mathcal{M}$ relation to vary between phases and depend strongly on how $\sigma_s^2$ and $\mathcal{M}$ are measured. The relation fails to capture the widths of the WNM and CNM density distributions, with possible contributing factors including non-isothermality and large-scale coherent motions. In contrast, we find the $\sigma_s^2-\mathcal{M}$ relation to tentatively hold for the log-normal portion of the H$_2$ distribution. The width of the CO density PDF is systematically overpredicted by the classical relation, resulting from the selective nature of CO as a molecular gas tracer.}
   {}


\maketitle

\nolinenumbers

\section{Introduction}

A central problem in astrophysics is understanding how the properties of molecular clouds influence the rate at which stars form. In this context, the density probability distribution function (PDF) of molecular clouds is of particular importance. The density PDF determines the fraction of gas above a critical threshold for gravitational collapse, and thus is a key ingredient in analytical models of star formation \citep{krumholz_general_2005, federrath_star_2012, burkhart_star_2018}. It is also closely tied to the physical properties of the cloud, including its turbulent and gravitational state \citep{passot_density_1998, kritsuk_density_2011, girichidis_evolution_2014}.

Early studies of the density PDF performed idealised numerical simulations, using an isothermal equation of state, periodic boundary conditions, and artificial driving mechanisms to invoke turbulent motions. The studies revealed the density PDF of an isothermal, supersonic medium to be well-approximated by a log-normal distribution \citep{Vazquez-Semadeni1994, padoan_universality_1997, scalo_density_1998},

\begin{equation}
    p(s) = \frac{1}{\sqrt{2\pi\sigma_s^2}} \exp\left(-\frac{(s - s_0)^2}{2\sigma_s^2}\right), 
    \label{eq:lognormal} 
\end{equation}

where $s = \text{ln}(\rho/\rho_0)$ is the logarithmic density contrast, and $s_0$ and $\sigma_s^2$ denote the mean and variance of the log-normal respectively. In the case of the volume-weighted density PDF, $s_0 = -\sigma_s^2/2$, due to the constraint of mass-conservation \citep{Vazquez-Semadeni1994}. Of particular significance was the finding that the width of the log-normal is controlled by turbulent properties of the medium, via the $\sigma_s^2-\mathcal{M}$ relation \citep{padoan_universality_1997, nordlund_density_1998, passot_density_1998}:

\begin{equation}
    \sigma_s^2 = \text{ln}(1+b^2\mathcal{M}^2). 
    \label{eq:isothermal_relation}
\end{equation}

Here, $b$ is the turbulent driving parameter and $\mathcal{M}$ is the sonic Mach number. The turbulent driving parameter describes the ratio of compressive to solenoidal turbulent modes, with $b\sim 1$ for purely compressive (i.e. curl-free) driving and $b\sim 1/3$ in the purely solenoidal (i.e. divergence-free) case \citep{federrath_density_2008, federrath_comparing_2010}. The emergence of the $\sigma_s^2-\mathcal{M}$ relation gave rise to `turbulence-regulated' analytical models of star formation, whereby the turbulent properties of a molecular cloud control the width of the log-normal density PDF, setting the fraction of gas at high density and thus the rate of star formation \citep{krumholz_general_2005, hennebelle_analytical_2011, padoan_star_2011, federrath_star_2012}.

In the case of an isothermal $\textit{gravo}$-turbulent medium, the shape of the density PDF is influenced by the self-gravity of the gas, in addition to its turbulent properties. At low densities, turbulence dominates over gravity and the density PDF retains its log-normal form. Beyond a critical density, however, a power-law tail emerges as a result of the gravitational collapse of the gas \citep{kritsuk_density_2011, girichidis_evolution_2014, burkhart_star_2018, Burkhart2019, jaupart_evolution_2020}. Observational studies of star-forming regions similarly report the column density PDF to have a multi-component structure, with a log-normal core and up to two power-law tails at high densities \citep{kainulainen_probing_2009, schneider_understanding_2015, veltchev_extraction_2019, schneider_understanding_2022, ma_gas_2022}. Building on this, \cite{burkhart_star_2018} extended the turbulence-regulated models of e.g. \cite{krumholz_general_2005} to the gravo-turbulent regime, deriving the star formation rate using a piecewise PDF, where the width of the log-normal core is again set by the $\sigma_s^2-\mathcal{M}$ relation.

Studies have further explored the density PDF and the $\sigma_s^2-\mathcal{M}$ relation in conditions beyond the isothermal gravoturbulent regime, with extensions of the classical relation accounting for the impact of magnetic pressure \citep{molina_density_2012}, varying adiabatic indices \citep{nolan_density_2015}, and polytropic equations of state \citep{federrath_density_2015}. The original $\sigma_s^2-\mathcal{M}$ relation for isothermal turbulence, however, remains the most widely used due to its simplicity and has been applied in a variety of contexts.
Observational studies frequently use the relation to infer the turbulent driving parameter $b$, combining measurements of the Mach number and the density PDF width. Studies have applied this method to CO isotopologue observations of star-forming regions \citep{brunt_density_2010, menon_turbulence_2020, Sharda2022} and to dust column density measurements combined with spectral line observations \citep{kainulainen_high-dynamic-range_2013, Federrath2016, kainulainen_relationship_2017}. 
Similarly, \cite{marchal_thermal_2021} and \cite{Gerrard2023} use HI emission observations to infer the turbulent driving parameter of the warm neutral medium (WNM). 
Alternatively, the relationship has been used to derive the Mach number, assuming a value for $b$. \cite{Berkhuijsen2008} use this approach to measure the Mach number of the diffuse ionised gas (DIG; also referred to as the warm ionised medium, WIM) in the solar neighbourhood. Studies also invoke the $\sigma_s^2-\mathcal{M}$ relation to connect cloud-scale properties to star formation activity, as in the recent work of  \cite{meidt_reconciling_2025}, who compare the Mach number and star formation efficiency of molecular clouds in the PHANGS-ALMA CO survey. 

The key assumption underpinning each of these studies is that the $\sigma_s^2-\mathcal{M}$ relation, empirically derived from isothermal simulations with artificially-driven turbulence, can be applied to the specific gas phase traced by the observational data (e.g. for highly-shielded molecular gas traced by CO). Whether this assumption holds for individual phases embedded within the complex, multi-phase ISM, however, has yet to be tested. In this paper, we address this question using the Cloud Factory, a suite of zoom-in simulations that model the detailed evolution of the multi-phase ISM and in which turbulence is self-consistently generated. This is achieved through the inclusion of time-dependent chemistry, a large-scale gravitational potential, self-gravity, star formation, and supernova feedback. This allows us to examine the validity of the $\sigma_s^2-\mathcal{M}$ relation across distinct ISM phases in a physically realistic environment. 

The plan for this paper is as follows: In Section~\ref{sec:sims}, we provide an overview of the Cloud Factory, before introducing the density PDFs of molecular cloud complexes extracted from the simulations in Section~\ref{sec:pdfs}. In Section~\ref{sec:defs}, we outline how $\sigma_s^2$, $b$, and $\mathcal{M}$ are measured for each ISM phase, before testing the $\sigma_s^2-\mathcal{M}$ relation in Section~\ref{sec:test}. We summarise our findings in Section~\ref{sec:conclusion}.

\section{Simulations}
\label{sec:sims}

\begin{figure*}
    \centering
    \includegraphics[width=0.9\linewidth]{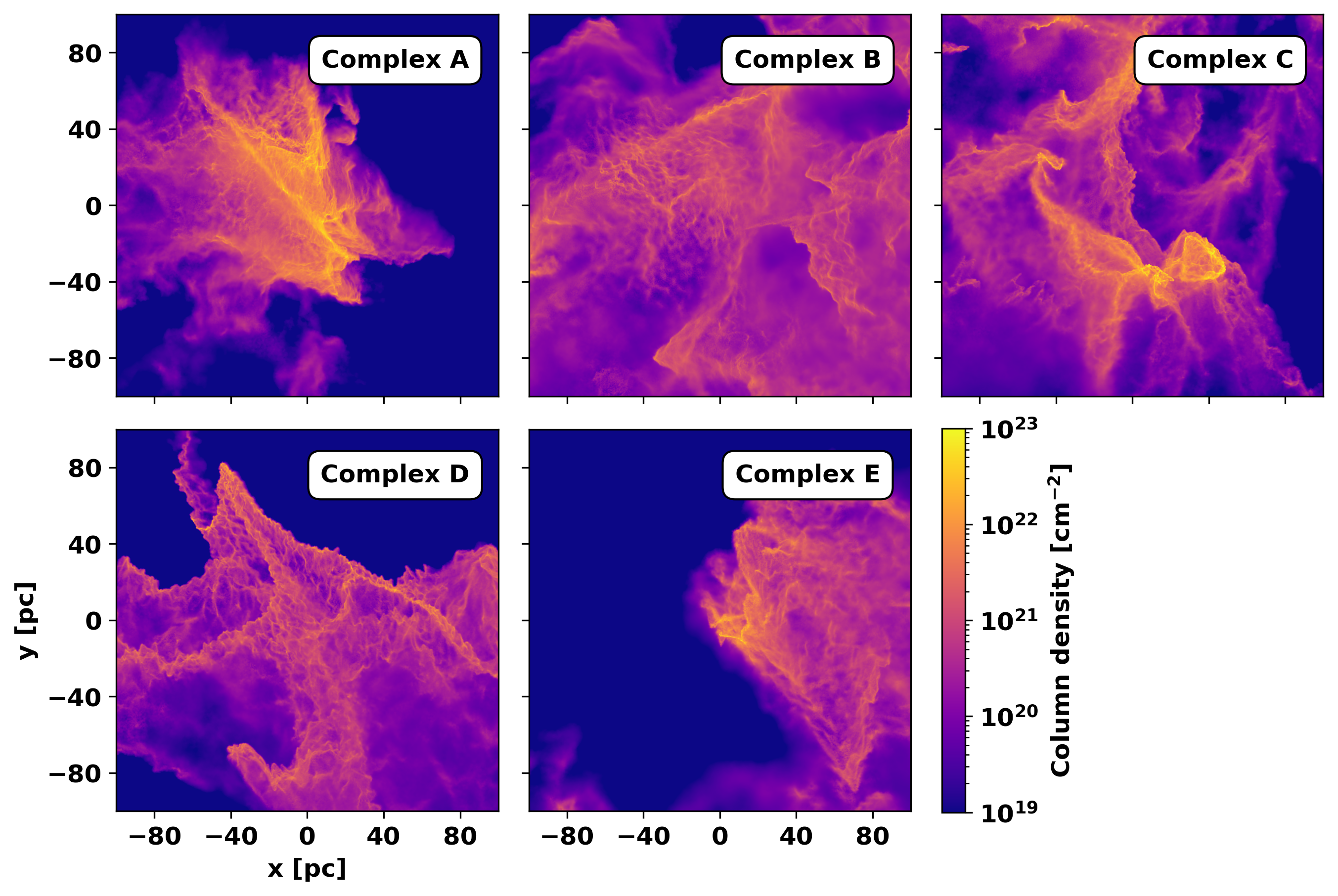}
    \caption{Total gas column density maps of the five molecular cloud complexes extracted from the Cloud Factory simulations. The complexes are viewed with the galactic disc face-on and at the middle evolutionary snapshot.}
    \label{fig:complexes}
\end{figure*}

The Cloud Factory is a suite of hydrodynamic simulations designed to study the evolution of the ISM at high resolution, whilst simultaneously capturing the large-scale environment in which it resides \citep{smith_cloud_2020, izquierdo_cloud_2020, feng_evolution_2024}. This is achieved through modelling a large-scale galactic disc and `zooming-in' to regions of interest. The simulations are performed in \textsc{Arepo}, which solves the (magneto) hydrodynamic (MHD) equations on an unstructured Voronoi mesh, with cells undergoing refinement and de-refinement in order to maintain a target mass \citep{springel_arepo_2010}. A modified version of the public release code is used, the key features of which are outlined below. For an in depth discussion of the simulations, see \cite{smith_cloud_2020} and \cite{feng_evolution_2024}.

\subsection{Galactic potential and self-gravity}

The Cloud Factory aims to simulate the behaviour of the ISM within a prototypical spiral galaxy. The gravitational influence of a dark matter halo and stellar component are modelled via a background potential, the analytical form of which is provided by \cite{mcmillan_potential_2017}, with a perturbation included to mimic the presence of spiral arms \citep{Smith2014}. The self-gravity of the gas is calculated using the \textsc{Arepo} gravitational tree \citep{springel_arepo_2010}.

\subsection{Chemical network}

The Cloud Factory simulations model the time-dependent evolution of hydrogen and CO, using the NL97 chemical network of \cite{glover_is_2012}. This employs the non-equilibrium hydrogen chemistry of \cite{glover&maclow2007a, glover&maclow2007b}, modelling the formation and dissociation of H$_{2}$, the ionisation of H, and the recombination of H$^{+}$. This is combined with the simplified treatment of CO introduced by \cite{nelson1997}, modelling its formation from C$^{+}$ and its destruction via photodissociation. The rate of photodissociation is calculated assuming a constant UV radiation field of solar neighbourhood strength \citep{draine1978}, with the impact of shielding 
solved for using the \textsc{treecol} algorithm of \cite{clark_treecol_2012}. The influence of radiative heating and cooling on the thermal state of the gas is determined using the atomic and molecular cooling function of \cite{clark_tracing_2019}.

\subsection{Star formation}

Sink particles are used within the Cloud Factory to represent sites of star formation. Sinks are non-gaseous, collisionless particles that can accrete from surrounding cells. A cell is a candidate for becoming a sink when the gas density, $\rho_\text{cell}$, exceeds a critical value, $\rho_\text{crit}$. In addition to $\rho_\text{cell} > \rho_\text{crit}$, a number of conditions must be fulfilled to ensure the gas within the region is collapsing, as outlined by \cite{bate1995} and \cite{federrath_sinks_2010}. At the level of refinement at which the simulations will be analysed in this study, the critical density for sink formation is $\rho_\text{crit} = 10^6$ cm$^{-3}$, with each sink roughly resembling a protostellar core. In line with typical star-forming efficiencies of protostellar cores, the stellar mass fraction of the sinks is set to 0.33 \citep{matzner2000}.

\subsection{Supernova feedback}

The Cloud Factory models feedback into the ISM from supernovae, whilst omitting early-stage stellar feedback such as photoionisation, radiation pressure and stellar winds. Two supernova components are modelled: 1) a component tied to the location of sink particles, with a supernova event triggered for each massive star contained within a sink; 2) a random component, where supernovae explode at random locations throughout the disc at a rate of one supernova every 300 yr \citep{tsujimoto1995}, mimicking Type 1a supernovae which will occur far from their natal environments. For a detailed discussion of the supernova feedback scheme employed, see \cite{tress_simulations_2021}.

\subsection{Zoom-in methodology}

Core to the Cloud Factory simulation methodology is the idea that the initial turbulent structure of molecular clouds is inherited from the large-scale galactic environment in which they reside. To this end, the Cloud Factory simulations undergo three phases, each with an increasing level of refinement. 

The first phase models the galactic disc with a target mass resolution of 1000 M$_\odot$, allowing the gas to reach a steady-state set by the background potential. After 150 Myrs, the resolution within a 3\,kpc box at a galactic radius of 8\,kpc is increased, with the target mass initially reducing to 100\,M$_\odot$ and later 10\,M$_\odot$. \cite{smith_cloud_2020} present two versions of this middle phase: a potential-dominated case, with no self-gravity and only random supernova feedback, and a feedback-dominated case, where self-gravity is turned on, sink particles can form, and both random and sink-related feedback is employed. For the purposes of this work, we focus on the feedback-dominated case. 

The middle phase of the simulation generates a range of structures within the high resolution box. The final phase `zooms-in' to regions of interest  identified within the box, in order to resolve their internal substructure. The mass outside a chosen region of radius $\sim200$\,pc is artificially reduced by a factor of $10^4$, leading to the de-refinement of the cells beyond the zoom-in volume. This makes it computationally feasible to reduce the target mass to 1\,M$_\odot$. The end result is a very high resolution view of molecular clouds which have been self-consistently generated and which can now be studied in detail.

\section{Density PDFs in the Cloud Factory}
\label{sec:pdfs}

Five zoom-in regions from the Cloud Factory simulations are studied in this work. A box of size 200$^3$\,pc is extracted from each region, centred on a molecular cloud complex of interest. The total gas column density maps of the selected molecular cloud complexes, viewed with the galactic disc face-on, are shown in Figure \ref{fig:complexes}. 

The complexes reside in a spiral arm of the galaxy and are chosen such that they have a diverse range of star-forming properties and are at an array of evolutionary stages: Complex A is a massive molecular cloud in the early stages of evolution, with no sinks initially present; Complex B is a diffuse region, devoid of star formation; Complex C comprises several molecular clouds and is an evolved state, with sinks present at the onset of the zoom-in phase; Complex D is a filamentary structure with limited star formation; and, finally, Complex E consists of a small molecular cloud surrounded by diffuse gas. Complexes A, C and E have previously been analysed in \cite{feng_evolution_2024}\footnote{Note that in \cite{feng_evolution_2024}, Complex E is labelled Region B.}, where the mass-to-length ratio of filaments was studied. We study the evolution of the molecular cloud complexes across time through selecting 7-8 `snapshots', output by the simulation every $10^5$ years and listed in Table~\ref{tab:snapshots}\footnote{While we study each molecular cloud complex at 7-8 evolutionary snapshots, the time periods over which they are studied varies due to the properties of each cloud. Complex B is studied over a limited time period as it is a short-lived feature that dissipates. Complex C has a large amount of sinks and thus is expensive to study, with the simulations also becoming increasingly unrealistic at later times due to the lack of early-stage feedback from newly-formed stars.}.

\begin{table}[]
    \centering
    \begin{tabular}{cc}
    \hline
    Complex & Snapshots \\
    \hline
     A & 253, 257, 262, 265, 267, 270, 272 \\
     B & 253, 254, 255, 256, 257, 258, 259 \\
     C & 253, 254, 255, 256, 257, 258, 259, 260 \\
     D & 253, 256, 258, 260, 262, 264, 267, 270 \\
     E & 253, 256, 258, 260, 262, 264, 266, 270 \\
     \hline
    \end{tabular}
    \caption{The \textsc{Arepo} snapshots analysed for each molecular cloud complex. Snapshots are output by \textsc{Arepo} every $10^5$ years, with the zoom-in phase starting at snapshot 249.}
    \label{tab:snapshots}
\end{table}

\begin{figure*}
    \centering
    \includegraphics[width=0.95\linewidth]{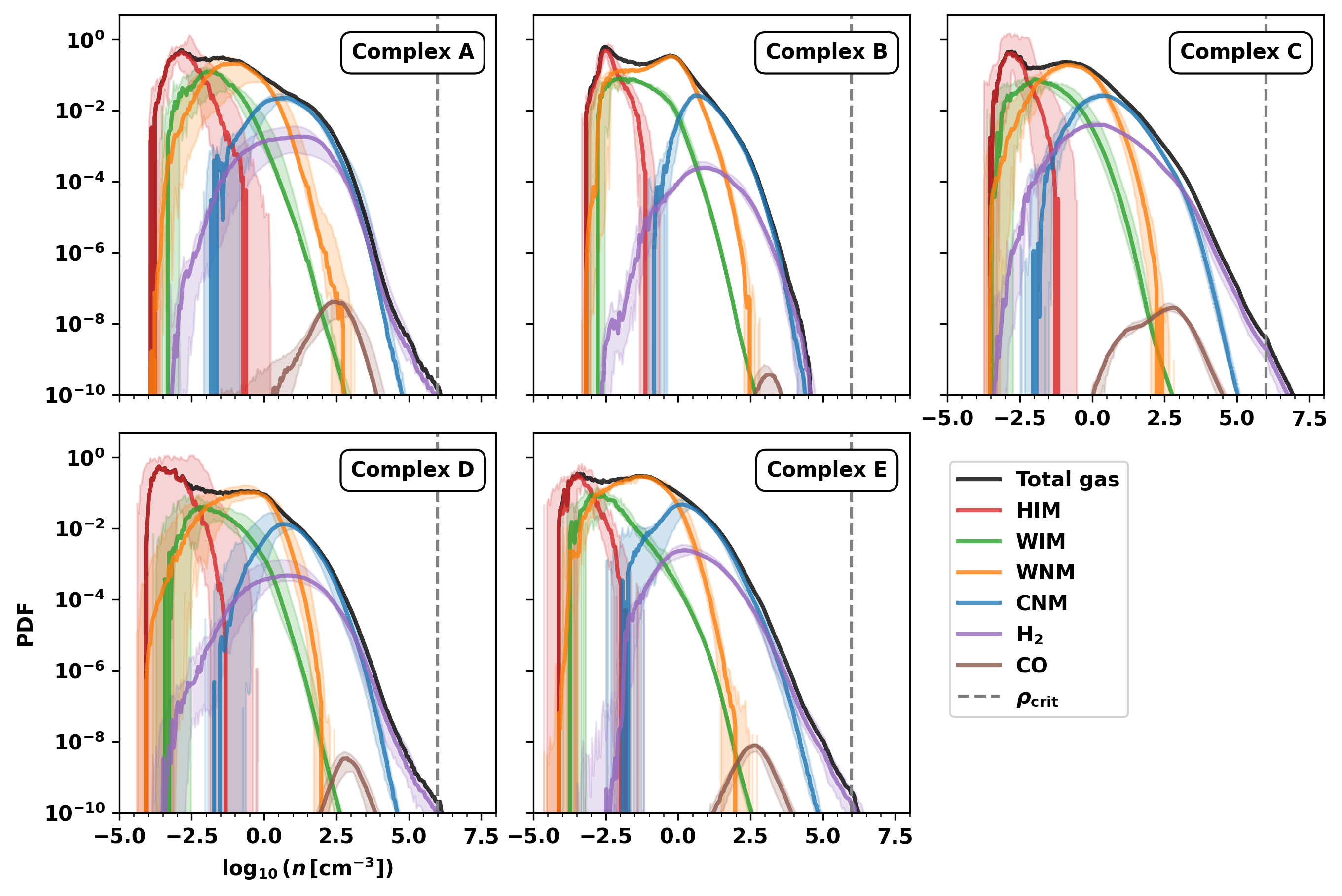}
    \caption{Density PDFs of the molecular cloud complexes from the Cloud Factory simulations. The solid lines delineate the time-averaged density PDFs, while the corresponding shaded areas illustrate the temporal variation in the PDFs over the studied time periods (see Table~\ref{tab:snapshots}). In black is the total density PDF, weighted by cell volume. Red, green, orange, blue, purple and brown correspond to the density PDFs of the HIM, WIM, WNM, CNM, H$_2$, and CO respectively, weighted by the product of the cell volume and the fractional abundance of the relevant species, and normalised by the total volume (see Table~\ref{tab:phases} for the definition of each phase). The critical density for sink creation, $\rho_\text{crit}=10^6$\,cm$^{-3}$, is shown with a dashed grey line.}
    \label{fig:density_pdfs}
\end{figure*}

\begin{figure*}
    \centering
    \includegraphics[width=0.95\linewidth]{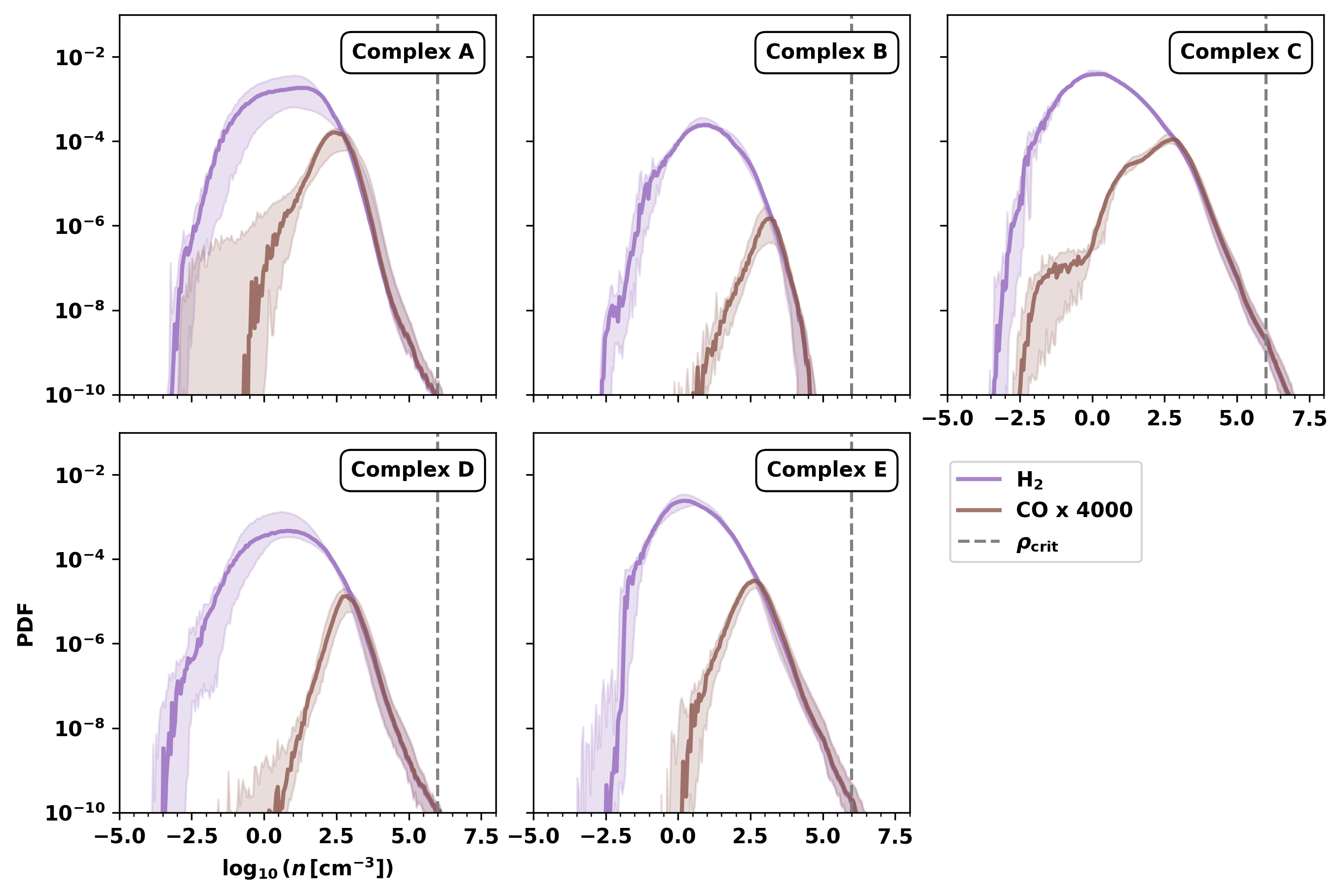}
    \caption{The molecular gas density PDFs shown in Figure~\ref{fig:density_pdfs}, with the weights of CO now scaled by a constant factor of 4000, for better comparison between the distributions of H$_2$ and CO. Again, solid lines delineate the time-averaged density PDFs, while the corresponding shaded areas illustrate the variation in the PDFs over the studied time periods. Purple and brown correspond to H$_2$ and CO respectively. The critical density for sink creation, $\rho_\text{crit}=10^6$\,cm$^{-3}$, is shown with a dashed grey line.}
    \label{fig:density_pdfs_CO_scaled}
\end{figure*}

Figure~\ref{fig:density_pdfs} presents the density PDFs of the five molecular cloud complexes, weighted by volume and averaged over all studied snapshots. Shown are the total density PDFs, as well as the contributions to the overall distributions from individual phases of the ISM. The phases are defined `thermochemically', with the criteria summarised in Table~\ref{tab:phases}, and are chosen to broadly align with the classically-defined phases of the ISM: ionised hydrogen is separated into hot and warm components at $10^{4.5}$\,K, corresponding to the hot and warm ionised medium (HIM and WIM); neutral hydrogen is split into warm and cold components at 500\,K, corresponding to the warm and cold neutral medium (WNM and CNM); and molecular hydrogen is treated as its own phase. We additionally study the density PDF of CO, which, while not classically constituting a distinct ISM phase, serves as a tracer of highly-shielded molecular gas. 

To compute the phase-specific density PDFs, each cell meeting the temperature criterion of the phase is weighted by the product of its volume and the fractional abundance of the relevant species:

\begin{equation}
    w_c = V_c\frac{n_{p,c}}{n_{{\rm tot},c}},
    \label{eq:weighting}
\end{equation}

where $c$ indexes individual cells, $V_c$ is the cell volume, $n_{p,c}$ is the number density of the species, and $n_{{\rm tot},c}$ is the total number density in the cell. The resulting density PDFs are found to be largely insensitive to the exact choice of temperature thresholds listed in Table~\ref{tab:phases}, as most gas lies well above or below them. In Figure~\ref{fig:density_pdfs}, the phase-specific density PDFs are then normalised by the total volume to show the contribution of the phase to the overall gas density distribution.

\begin{table}[]
    \centering
    \begin{tabular}{llc}
    \hline
    Phase & Species & Temperature \\
    \hline 
    HIM & H$^+$ & $T \geq 10^{4.5}\,$K \\
    WIM & H$^+$ & $T < 10^{4.5}\,$K \\
    WNM & H & $T \geq 500\,$K \\
    CNM & H & $T < 500\,$K \\
    H$_2$ & H$_2$ & -- \\
    CO & CO & -- \\
    \hline
    \end{tabular}
    \caption{The thermochemically-defined ISM phases studied in this work, along with the relevant species and temperature thresholds.}
    \label{tab:phases}
\end{table}

A first inspection of Figure~\ref{fig:density_pdfs} reveals the total density PDFs of each complex to deviate significantly from the simple log-normal expected for an isothermal medium. This is unsurprising given the multi-phase ISM generated by the Cloud Factory, with gas temperatures ranging from $\sim10 - 10^6$\,K. The total density PDFs are instead multi-modal and, in all but Complex B, exhibit power-law tails at high density. 

The phase-specific PDFs provide insight into this multi-modal structure. Ionised hydrogen dominates the density PDF at low densities, with the HIM responsible for the narrow peak at $\sim0.001$ cm$^{-3}$
and the WIM occupying intermediate densities between the hot ionised and warm neutral gas. The absence of early-stage stellar feedback in the Cloud Factory likely means H$^+$ is underestimated in the simulations and, as such, we limit our discussion of the HIM and WIM.

At intermediate densities between $\sim0.01$ and 1000 cm$^{-3}$, neutral hydrogen accounts for the majority of the gas, with the density distributions of the WNM and CNM overlapping significantly. This is in contrast to classical two-phase models of HI, where thermal instability leads to strong segregation of the two phases and distinct peaks in the density PDF \citep{Field1965, Mckee1977, audit_thermal_2005}. The overlap arises due to turbulent mixing, which allows gas to exist at intermediate, classically-forbidden temperatures \citep{hennebelle_structure_2007, Vazquez2009}. 
The density PDF of the WNM deviates from log-normality at low densities where a significant fraction of the gas is ionised, 
most notably in Complex B.
In contrast, the density distributions of the CNM appear approximately log-normal.

Beyond $\sim1000$ cm$^{-3}$, molecular hydrogen dominates the density PDF. Interestingly, the HI-to-H$_2$ transition roughly coincides with the density at which the power-law tail emerges, as also noted in the observational study of \cite{burkhart_lognormal_2015}. The quiescent Complex B exhibits no power law tail, in agreement with observations of column density PDFs, which report power-law tails only in star-forming regions \citep{kainulainen_probing_2009, ma_gas_2022}. Complexes A and D, in contrast, show evidence for two power-laws of differing slopes, with a possible second, shallower tail emerging at $\sim10^{4.5}$ cm$^{-3}$. 

Multiple power-law tails have been reported in the column density PDFs of molecular clouds \citep{Schneider_two_tails_2015, Stanchev2015, Pokhrel2016, schneider_understanding_2022}, as well as in analytical and numerical studies \citep{kritsuk_density_2011, Murray2017, jaupart_evolution_2020, khullar_density_2021}. While the first power-law tail is generally understood to mark the onset of self-gravitation, the origin of the second tail remains debated. Proposed explanations include rotational support from collapsing cores, increased thermal pressure, or later stages of free-fall collapse \citep{schneider_understanding_2015, jaupart_evolution_2020, khullar_density_2021}. Our ability to study the possible second tail is limited by the sink creation density threshold, $10^{6}$ cm$^{-3}$, and we leave a detailed study of the power-law tail(s) and the connection to star formation to a future study.

Finally, we examine the density PDF of CO and compare it to that of H$_2$. Due to the low abundance of CO relative to H$_2$, direct comparison between the distributions is difficult from Figure~\ref{fig:density_pdfs} alone. As such, we remake Figure~\ref{fig:density_pdfs}, including now only H$_2$ and CO and rescaling the CO distribution by a constant factor of 4000. The resulting PDFs are shown in Figure~\ref{fig:density_pdfs_CO_scaled}. We can now clearly see the density PDF of CO closely follows the distribution of H$_2$ at high densities, exhibiting a power-law tail of the same slope. 
At densities below $\sim1000$ cm$^{-3}$, however, CO is no longer a reliable tracer of the H$_2$ distribution, with a much narrower central component. This is the result of CO photodissociating in weakly-shielded regions. 

\section{$\sigma_s^2$, $b$, and $\mathcal{M}$ of individual ISM phases}
\label{sec:defs}

In order to test the $\sigma_s^2-\mathcal{M}$ relation in individual phases of the ISM, we must first measure $\sigma_s^2$, $b$, and $\mathcal{M}$. Here we outline our definitions of each parameter, which are non-trivial when moving beyond the simple isothermal regime.

\subsection{The density variance, $\sigma_s^2$}
\label{sec:width}

\begin{figure}
    \centering
    \includegraphics[width=0.9\linewidth]{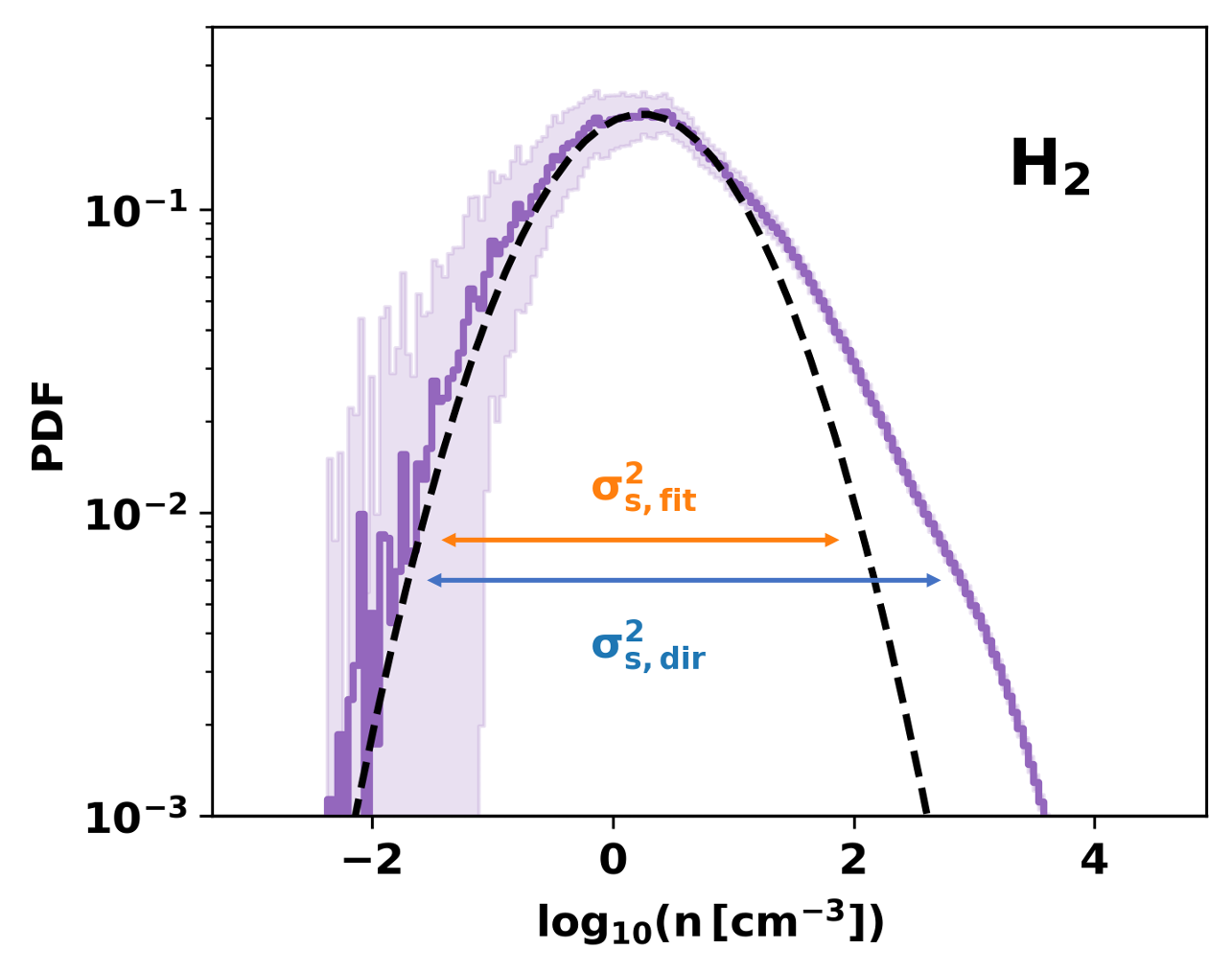}
    \caption{The density PDF of H$_2$ for Complex C at the middle evolutionary snapshot. The purple line is the density PDF, while the purple shaded area is the bootstrapped error on each PDF bin. Note that it is now not the temporal variation in the density PDF. The best-fitting log-normal model is overlaid as a black dashed line. The two arrows illustrate the difference in the parameters $\sigma_{\rm s,fit}^2$ and $\sigma_{\rm s,dir}^2$, with $\sigma_{\rm s,fit}^2$ measuring the variance of the log-normal component and $\sigma_{\rm s,dir}^2$ the variance of the whole distribution.}
    \label{fig:h2_dir_v_fit}
\end{figure}

\begin{figure*}
    \centering
    \includegraphics[width=0.95\linewidth]{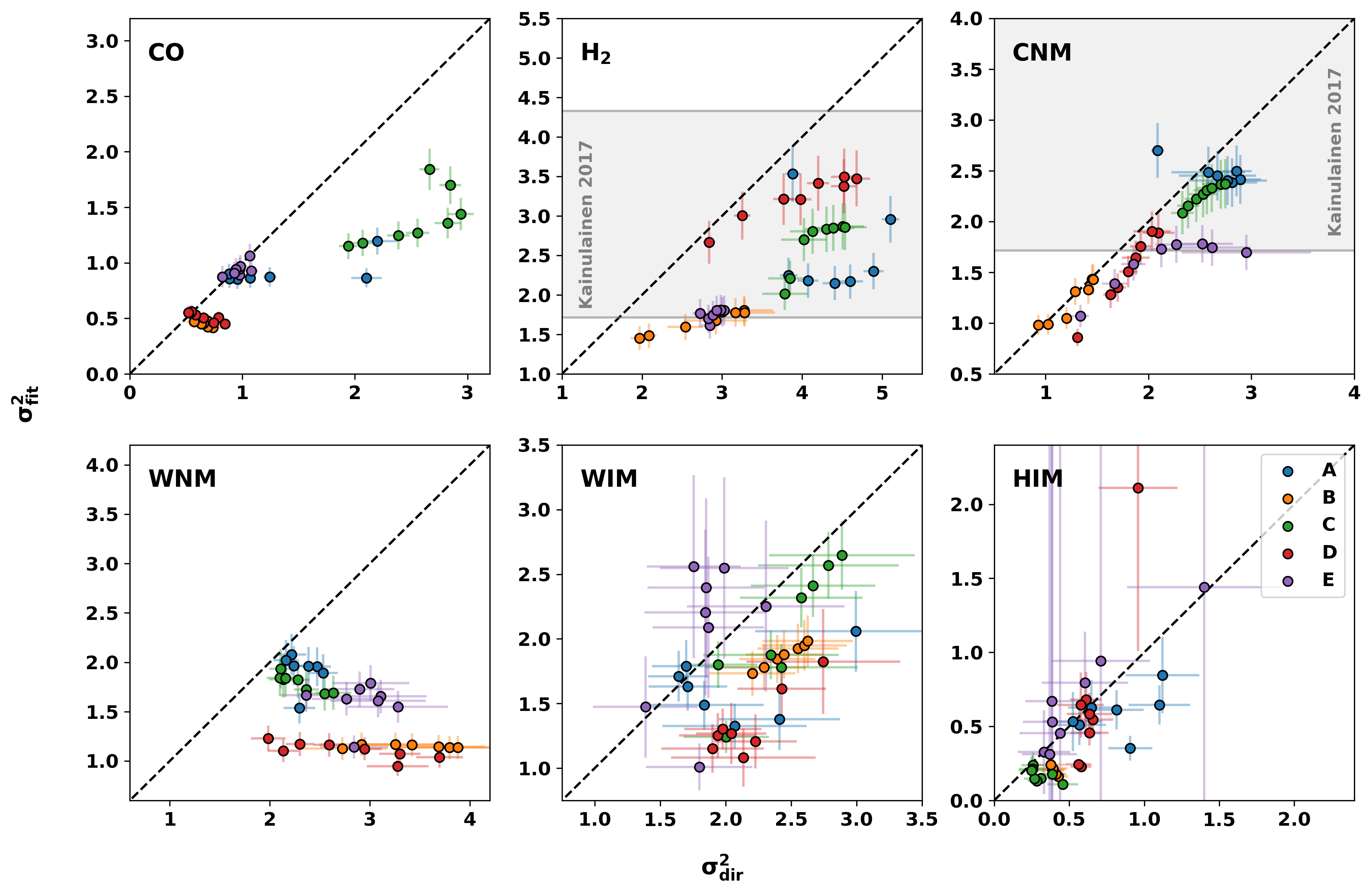}
    \caption{The width of the phase-specific density PDFs measured via fitting a log-normal function to the density PDF ($\sigma_{\rm s,fit}^2$) versus direct calculation over the entirety of the PDF ($\sigma_{\rm s,dir}^2$). The black dashed line delineates $\sigma_{\rm s,fit}^2 = \sigma_{\rm s,dir}^2$. The coloured markers show the values for each molecular cloud complex at each evolutionary snapshots over which they are studied. The grey horizontal bands overplotted in the H$_2$ and CO panels show the range of density PDF widths reported in Table 1 of \cite{kainulainen_relationship_2017}.}
    \label{fig:dir_v_fit}
\end{figure*}

The variance of the density PDF of each phase can be characterised in two different ways: 1) $\sigma_{\rm s,dir}^2$, accounting for the total variance of the PDF, and 2) $\sigma_{\rm s,fit}^2$, measuring only the variance of the approximately log-normal component of the density distribution. 

To compute $\sigma_{\rm s,dir}^2$ of each phase, we directly calculate the variance of the logarithmic density, $s$, weighting each cell, $c$, by $w_c$, i.e. the product of the cell volume and the fractional abundance of the phase (see Equation~\ref{eq:weighting}):

\begin{equation}
\sigma_\text{s,dir}^2 =
\frac{\sum_c w_c(s_c - \langle s\rangle_w)^2}{\sum_c w_c};
\end{equation}

\begin{equation}
\langle s\rangle_w =
\frac{\sum_c w_c s_c}{\sum_c w_c}.
\end{equation}

Given that \textsc{Arepo} outputs no uncertainties on cell properties, we estimate the error on $\sigma_\text{s,dir}^2$ via bootstrapping: we randomly sample 1$\%$\footnote{We use 1$\%$ of the total number of cells for computational efficiency, as each simulation box contains $\sim10^6$ cells.} of the total number of cells with replacement 500 times and take the standard deviation of the resulting $\sigma_\text{s,dir}^2$ distribution as the error.

To measure $\sigma_\text{s,fit}^2$, we fit a log-normal function (Equation~\ref{eq:lognormal}) to each phase-specific density PDF, restricting the fit to the density range where the distribution appears approximately log-normal. In this way, we isolate the ‘turbulence-driven’ variance, excluding the impact of, for example, the high density power-law tail arising from gravitational collapse. 

The fit is performed using an MCMC routine, implemented with the Python package \textsc{emcee} \citep{Foreman2013} and employing 50 walkers, a burn-in run of 1000 iterations, and a production run of 5000 iterations. $\sigma_\text{s,fit}^2$ is the single free parameter of the model, for which we impose a positive uniform prior, as $s_0$ is constrained by $s_0= -\sigma_\text{s,fit}^2/2$. Errors on each PDF bin are estimated again via bootstrapping and adopted for use in the likelihood function.
Convergence of the MCMC is confirmed via visual inspection of the trace plots and through requiring a Gelman-Rubin statistic of $\leq1.1$. The best-fitting value of $\sigma_\text{s,fit}^2$ is chosen such that it maximises the log-posterior, with the corresponding error estimated from the 16th and 84th percentiles of the posterior distribution. To account for the uncertainty associated with the choice of density range over which the fit is performed, we add in quadrature a $10\%$ fractional error to the posterior-derived uncertainty.

To illustrate the difference between $\sigma_\text{s,dir}^2$ and $\sigma_\text{s,fit}^2$, in Figure~\ref{fig:h2_dir_v_fit} we present the H$_2$ density PDF for Complex C at the middle evolutionary snapshot, with the fitted log-normal overlaid. We can clearly see that $\sigma_\text{s,fit}^2$ measures the variance of the central log-normal component, whereas $\sigma_\text{s,dir}^2$ includes the contribution from the power-law tail, leading to $\sigma_\text{s,fit}^2 \lesssim \sigma_\text{s,dir}^2$. 

We compare $\sigma_\text{s,dir}^2$ and $\sigma_\text{s,fit}^2$ for each phase in Figure~\ref{fig:dir_v_fit}. We similarly find $\sigma_\text{s,fit}^2 \lesssim \sigma_\text{s,dir}^2$ for CO, with the offset now arising due to the combined effect of the high-density power-law tail and deviations from log-normality at low densities (see Figure~\ref{fig:density_pdfs_CO_scaled}). Diffuse excess gas at low density again leads to $\sigma_\text{s,fit}^2 \lesssim \sigma_\text{s,dir}^2$ for the WNM, while for the CNM, WIM, and HIM, we find $\sigma_\text{s,fit}\approx\sigma_\text{s,dir}$, consistent with the distributions being close to log-normal. Discrepancies may still arise, however, as a result of turbulent intermittency, which enhances the tails of the distributions and thus can inflate $\sigma^2_\text{s,dir}$ relative to $\sigma_\text{s,fit}^2$ \citep{federrath_comparing_2010, hopkins_model_2013}.

It is interesting to compare our measured density PDF widths to those reported by observational studies. Observations probe the column density PDF, from which they must then infer the volume density PDF. \cite{Kainulainen2014} achieve this through modelling the structure of column density maps using an ensemble of prolate spheres. \cite{kainulainen_relationship_2017} then present a compilation of density PDF widths recovered using this approach (their Table~1), to which we compare to here. 

The density PDFs widths reported in \cite{kainulainen_relationship_2017} are derived from near-infrared extinction and \textit{Herschel} dust emission data. As dust is generally accepted to trace both cold atomic and molecular gas, we make comparison to the widths of the H$_2$ and CNM density PDFs. The observationally-derived widths are obtained by fitting log-normal functions to the recovered volume density PDFs and thus correspond most closely to our definition of $\sigma_\text{s,fit}^2$. We therefore plot the minimum and maximum widths reported by \cite{kainulainen_relationship_2017} as horizontal lines in the H$_2$ and CNM panels of Figure~\ref{fig:dir_v_fit}. We find the observationally-inferred range aligns well with our measurements of the H$_2$ density distributions, while our measured CNM widths often fall below the observed range.

\subsection{The turbulent driving parameter, $b$}
\label{sec:b}

\begin{figure}
    \centering
    \includegraphics[width=0.9\linewidth]{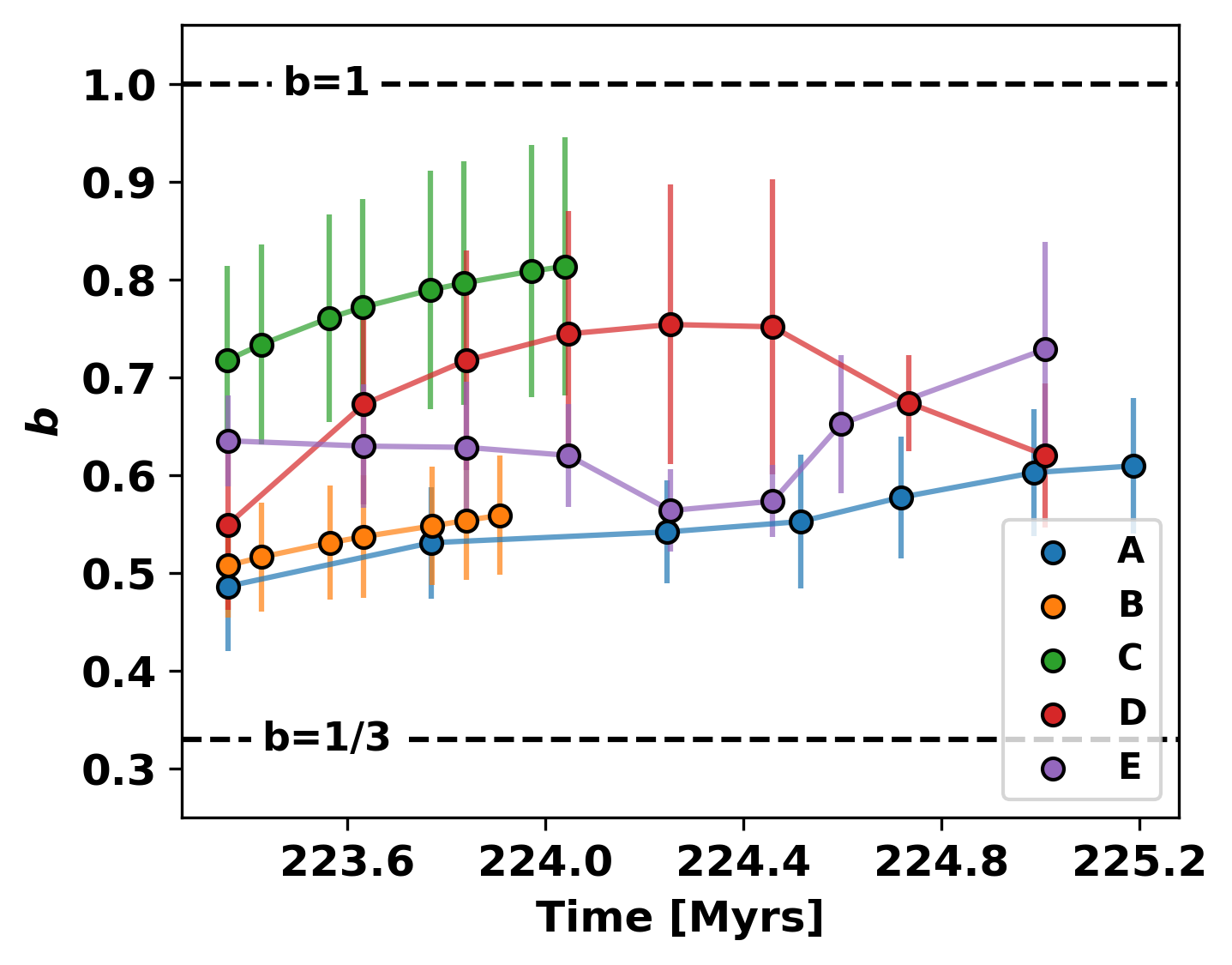}
    \caption{The turbulent driving parameter as a function of time, for each of the molecular cloud complexes extracted from the Cloud Factory. The black dashed lines delineate the expected value of $b$ for compressive ($b=1$) and solenoidal ($b=1/3$) driving. The error bars show the standard deviation of $b(k)$ across all studied scales.}
    \label{fig:b_evolution}
\end{figure}

To measure $b$, the turbulent driving parameter, independent of assuming the $\sigma_s^2-\mathcal{M}$ relation holds, we use the method introduced by \cite{pan_supernova_2016}. \cite{pan_supernova_2016} argue $b$ can be estimated from the compressive ratio, $\chi$, of the velocity field:

\begin{equation}
    b = \sqrt{\frac{\chi}{\chi + 1}},
\end{equation}

\begin{equation}
    \label{eq:chi}
    \chi = \frac{\langle |\text{\textbf{v}}_\mathrm{c}(\mathbf{x})|^2 \rangle}{\langle |\text{\textbf{v}}_\mathrm{s}(\mathbf{x})|^2 \rangle}.
\end{equation}

Here, $\textbf{v}_c(\textbf{x})$ and $\textbf{v}_s(\textbf{x})$ are the compressive (i.e. $\nabla \times \textbf{v}_c = 0$) and solenoidal (i.e. $\nabla \cdot \textbf{v}_s = 0$) components of the velocity field $\textbf{v}(\textbf{x})$, extracted via Helmholtz decomposition. The decomposition is performed using Fast Fourier Transforms (FFTs), which require uniformly-spaced data. Because the native \textsc{Arepo} data is mapped onto an unstructured Voronoi mesh, we first interpolate the velocity data onto regular Cartesian grids of resolution $512^3$ ($\Delta=0.39$\,pc) using the Python package \textsc{yt} \citep{Turk2011}. To reduce artefacts from the non-periodicity of the simulation boxes, we apply a Hanning window to the interpolated velocity fields, tapering the amplitude to zero over 5 cells at each boundary. In Appendix~\ref{sec:appendix_helmholtz}, we test the sensitivity of the results to the interpolation resolution and the window size, and find that the inferred $b$ is insensitive to both of these choices.

The  time-evolution of $b$ for each molecular cloud complex is displayed in Figure~\ref{fig:b_evolution}. $b$ ranges from $\sim0.5-0.8$, indicating that a mixture of compressive and solenoidal turbulent modes are driven in the simulation, and varies across time and between complexes. Similar values for the turbulent driving parameter have been reported in the simulations of \cite{kortgen_turbulence_2020} and \cite{Gerrard2025}. 

It is important to consider whether a single global $b$ is representative of the turbulent driving within individual ISM phases, which occupy different spatial scales. To explore this, we introduce the scale-dependent turbulent driving parameter,  $b(k)$, expressed in terms of the scale-dependent compressive ratio, $\chi(k)$:

\begin{equation}
    b_\chi(k) = \sqrt{\frac{\chi(k)}{\chi(k) + 1}}.
\end{equation}

Equation~\ref{eq:chi} could have been equivalently formulated in Fourier space, i.e. $\chi = \langle |\bm{\tilde{\textbf{v}}}_c(\bm{k})|^2\rangle/\langle |\bm{\tilde{\textbf{v}}}_s(\bm{k})|^2 \rangle$
where $\bm{\tilde{\textbf{v}}}_c(\bm{k})$ and $\bm{\tilde{\textbf{v}}}_s(\bm{k})$ are the Fourier-transformed compressive and solenoidal velocity components and the average is performed over all wavevectors $\bm{k}$. If instead of averaging over all $\bm{k}$, the average is taken over spherical shells of radius $k = |\bm{k}|$, the scale-dependence of the compressive ratio is preserved. In this case, $\chi(k)$ is simply the ratio of the 1D power spectra of the compressive and solenoidal velocity components:

\begin{equation}
    \chi(k) = \frac{\langle|\bm{\tilde{\textbf{v}}}_c(\bm{k})|^2\rangle_k}{\langle|\bm{\tilde{\textbf{v}}}_s(\bm{k})|^2\rangle_k} = \frac{P_{c}(k)}{P_{s}(k)}.
\end{equation}

\begin{figure}
    \centering
    \includegraphics[width=0.85\linewidth]{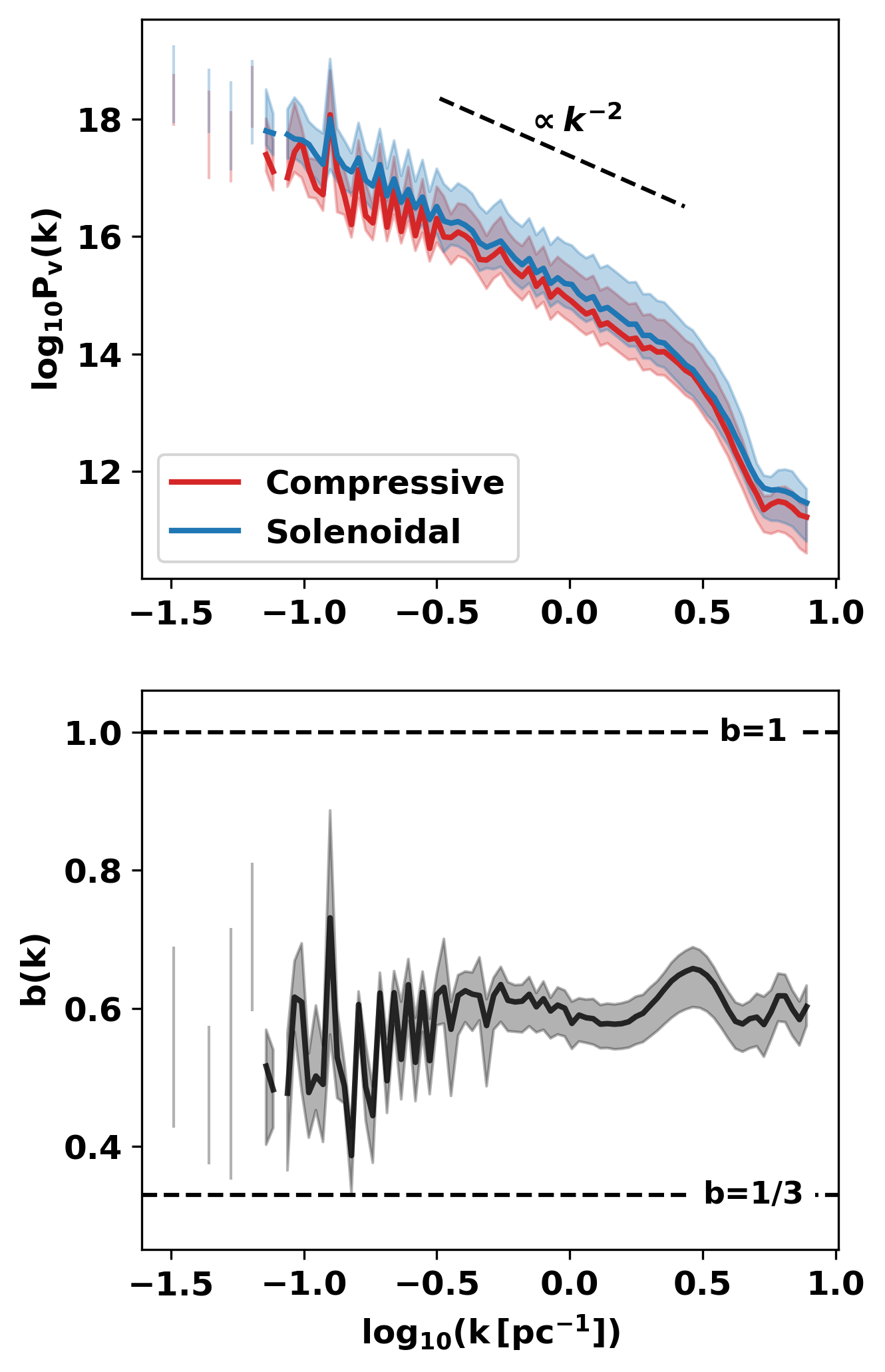}
    \caption{\textit{Upper:} The time-averaged 1D power spectra of the compressive (red line) and solenoidal (blue line) velocity components of Complex A, with corresponding shaded areas illustrating the variation over the studied time period. The spectra are plotted up to the Nyquist frequency. The black dashed line shows the expected slope of the velocity power spectrum for Burgers turbulence. \textit{Lower:} The time-averaged $b(k)$ as a function of the 1D wavenumber, $k$, again for Complex A. The shaded area shows again the variation over time. The black dashed lines delineate the expected value of $b$ for compressive ($b=1$) and solenoidal ($b=1/3$) driving.}
    \label{fig:scale_dependence_b}
\end{figure}

Figure~\ref{fig:scale_dependence_b} presents the time-averaged 1D power spectra of the compressive and solenoidal velocity fields, as well as the resulting scale-dependence of $b(k)$, for Complex A. The power spectra of the compressive and solenoidal velocity components are similar in amplitude and slope, with the spectra following the expected scaling relationship for supersonic, compressible turbulence, $P_v(k)\propto k^{-2}$ \citep{Burgers1984, bec2007}. The spectra exhibit a downturn at high-$k$, which is often attributed to numerical dissipation, indicating the scale at which turbulence is no longer reliably resolved in the simulations (\citealp{federrath_universality_2013, Padoan2016, kobayashi_nature_2022}). Here, however, this downturn is the result of the interpolation and windowing procedure, as is shown in Appendix~\ref{sec:appendix_helmholtz}. 

The similar behaviour of the compressive and solenoidal velocity spectra is reflected in the behaviour of $b(k)$, which consequently shows little variation across scales. This is perhaps unsurprising given the primary drivers of turbulence in the Cloud Factory are supernovae and large-scale galactic dynamics, which are thought to inject turbulence on scales of the order of $\gtrsim100$\,pc \citep{Wada2002, deAvillez2007, Falceta2015, Chamandy2020}. If the simulations included turbulent driving mechanisms with smaller injection scales, e.g. protostellar jets and outflows, $b(k)$ may show greater variation.  

Given the roughly constant values of $b(k)$ across scales, similarly observed for the remaining complexes, we conclude the globally-calculated $b$ is a reasonable measure of the turbulent driving parameter in each ISM phase. We estimate the uncertainty on $b$ as the standard deviation of $b(k)$ over all wavenumbers to account for possible variation across scales.

\subsection{The Mach number, $\mathcal{M}$}
\label{sec:mach}

\begin{figure*}
    \centering
    \includegraphics[width=0.95\linewidth]{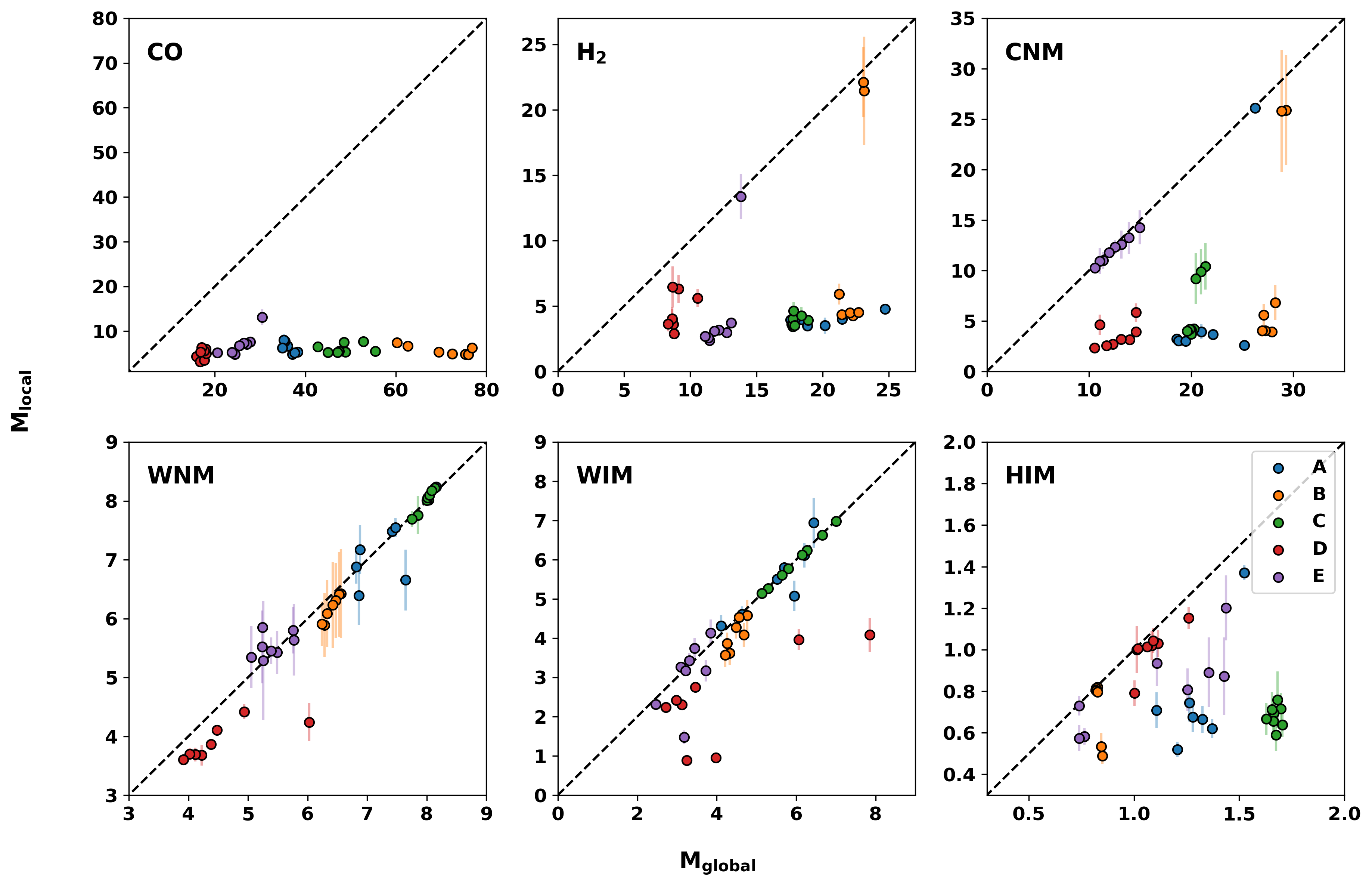}
    \caption{The local Mach number ($\mathcal{M}_{\rm{local}}$) versus the global Mach number ($\mathcal{M}_{\rm{global}}$) for each phase. The black dashed line delineates $\mathcal{M}_\text{local} = \mathcal{M}_\text{global}$. The coloured markers show the calculated values for each molecular cloud complex at each evolutionary snapshots over which they are studied.}
    \label{fig:mach_global_v_local}
\end{figure*}

\begin{figure}
    \centering
    \includegraphics[width=\linewidth]{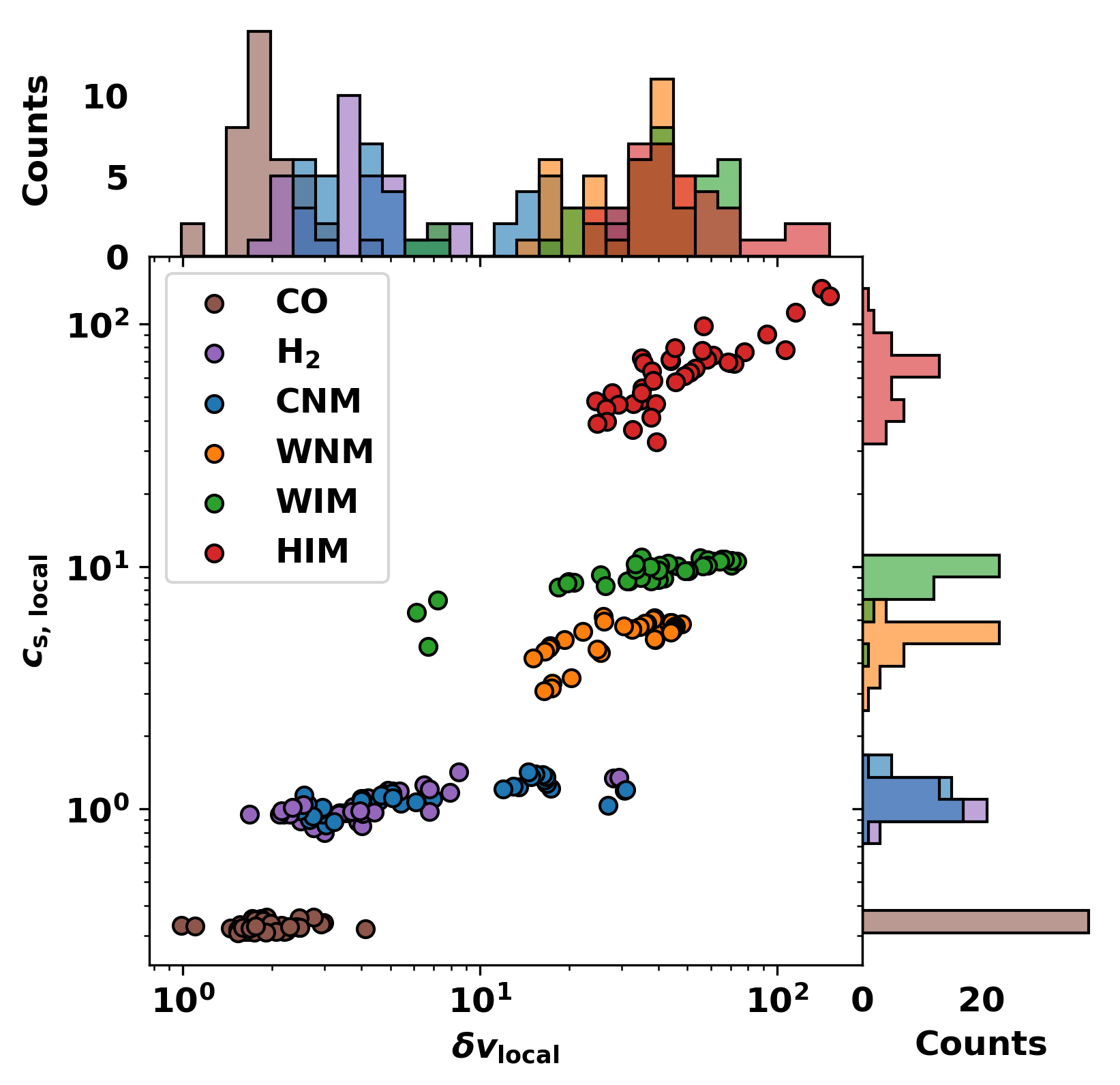}
    \caption{Scatter plot of the local sound speed ($c_\text{s, local}$) versus the local velocity dispersion ($\delta v_\text{local}$) for each phase. Accompanying histograms show the distribution of $c_\text{s, local}$ (right) and $\delta v_\text{local}$ (above).}
    \label{fig:cs_v_vel_disp}
\end{figure}

The Mach number for an isothermal medium is straightforwardly defined as $\mathcal{M} = \delta v/c_s$, where $\delta v$ is the volume-weighted velocity dispersion and $c_s$ is the uniform sound speed. Extending this definition to individual phases of the ISM, one might define a phase-specific Mach number by again weighting each cell by $w_c$ (see Equation~\ref{eq:weighting}):

\begin{equation}
    \label{eq:mach_number}
     \mathcal{M} = \frac{\delta v_{3\text{D},w}}{\langle c_{s} \rangle_w}.
\end{equation}

Here, $\delta v_{3\text{D},w}$ is the weighted 3D velocity dispersion,

\begin{equation}
    \label{eq:3d_vel_disp}
    \delta v_{3\text{D},w} = \sqrt{\delta v^2_{x,w} + \delta v^2_{y,w} + \delta v^2_{z,w}} ,
\end{equation}

dependent on the weighted 1D velocity dispersions,

\begin{equation}
\label{eq:vel_disp}
       \delta v_{i,w} = \sqrt{\frac{\sum_cw_{c}(v_{i,c} - \langle v_i \rangle_{w})^2}{\sum_c w_{ c}}};
\end{equation}

\begin{equation}
\langle v_i\rangle_w =
\frac{\sum_c w_c v_{i,c}}{\sum_c w_c}.
\end{equation}

The weighted average sound speed is

\begin{equation}
\label{eq:avg_cs}
\langle c_{s}\rangle = \frac{\sum_c w_{c}\, c_{s,c}} {\sum_c w_{\text{c}}};
\end{equation}

\begin{equation}
\label{eq:cs}
    c_{s,c} = \sqrt{\frac{k_B T_c}{\mu_c m_p}},
\end{equation}

where $T_c$ and $\mu_c$ are the cell temperature and mean molecular weight respectively.

We will refer to this as the phase-specific \textit{global} Mach number, $\mathcal{M}_\text{global}$, as it is calculated over the entirety of each simulation box. The extent to which this global Mach number reflects the internal turbulence of the phase depends on its spatial distribution, as noted by \cite{kobayashi_nature_2022}. For highly structured phases, such as the shielded molecular phase traced by CO, dense regions are embedded within more diffuse gas. In such cases, the velocity dispersion computed by 
Equation~\ref{eq:vel_disp} is dominated by the relative motions between regions, rather than capturing their internal turbulence. As a result, the Mach number is artificially inflated.

To obtain a Mach number more representative of the turbulent motions of each phase, we first identify spatially distinct regions within that phase. This is performed using \textsc{Hdbscan} \citep{campello2013density}, a hierarchical clustering algorithm which we apply to all cells whose abundance exceeds a phase-specific threshold. This threshold is chosen such that 99\% of the phase-specific density PDF is accounted for above it. \textsc{Hdbscan} requires a user-chosen minimum cluster size, setting the minimum number of cells necessary to be identified as a distinct structure. We tested minimum cluster sizes of 4, 8, 16, and 32 cells and found the clustering results largely insensitive to the exact value. We adopt 8 cells as the minimum cluster size going forwards.

For each identified cluster, we calculate the Mach number using Equations~\ref{eq:mach_number}-\ref{eq:cs} as before. Finally, we average over all clusters to obtain the phase-specific \textit{local} Mach number, $\mathcal{M}_\text{local}$, weighting by the volume of each cluster. The uncertainty on $\mathcal{M}_\text{local}$ is estimated as the standard error of the distribution of cluster Mach numbers. 

The global and local Mach numbers are computed for each phase and compared in Figure~\ref{fig:mach_global_v_local}. As expected, $\mathcal{M}_\text{global}$ and $\mathcal{M}_\text{local}$ differ significantly for CO: $\mathcal{M}_\text{global}$ $\sim20-80$ while $\mathcal{M}_\text{local}$ spans the more plausible range $\sim3-8$, in line with observational studies \citep{brunt_density_2010, kainulainen_high-dynamic-range_2013, orkisz_turbulence_2017}. In contrast, $\mathcal{M}_\text{global} \approx \mathcal{M}_\text{local}$ for the WNM and the WIM, reflecting their relatively smooth spatial distributions. The picture is more complicated for H$_2$, the CNM and the HIM, with varying levels of agreement between $\mathcal{M}_\text{global}$ and $\mathcal{M}_\text{local}$. We attribute this to differing morphologies between the molecular cloud complexes and as they evolve over time.

To gain further insight into the local Mach numbers of each phase, we examine the average velocity dispersions and sound speeds of the clusters identified by \textsc{Hdbscan}. We similarly denote these quantities $\delta v_\text{local}$ and $c_{s,\text{local}}$, and plot their distributions in Figure~\ref{fig:cs_v_vel_disp}. Within each phase we observe a wide range of velocity dispersions, covering up to an order of magnitude. In contrast, each phase has a relatively well-defined sound speed, indicating $c_{s,\text{local}}$ is largely independent of cloud-specific properties. Consequently, variations in the phase-specific Mach number are primarily the result of variations in velocity dispersion. 

\section{Testing the $\sigma_s^2-\mathcal{M}$ relation} 
\label{sec:test}

With $\sigma_s^2$, $b$, and $\mathcal{M}$ measured, we can now assess in which phases, if any, the classical $\sigma_s^2-\mathcal{M}$ relation is applicable. For each phase, we calculate the theoretically-expected density variance ($\sigma^2_\text{s,theory}$), using Equation~\ref{eq:isothermal_relation} and inputting the turbulent driving parameter found via Helmholtz decomposition, $b$, and the local Mach number, $\mathcal{M}_\text{local}$.

We will now compare $\sigma^2_\text{s,theory}$ to the density variance measured from the simulations for each phase, obtained via direct measurement ($\sigma^2_\text{dir}$) and via log-normal fitting ($\sigma^2_\text{fit}$). To quantify the relationship between the measured and predicted values, we use two metrics: 

\begin{enumerate}
    \item $\chi^2_{y=x}$: The reduced $\chi^2$ of the null hypothesis $\sigma^2_\text{dir}$ (or $\sigma^2_\text{fit}$) = $\sigma^2_\text{s,theory}$,  weighted by the combined uncertainties on both parameters. This metric quantifies the level of agreement between the measured and predicted values.
    \item $\chi^2_{\text{ODR}}$: The reduced $\chi^2$ of an orthogonal distance regression (ODR) linear fit between $\sigma^2_\text{dir}$ (or $\sigma^2_\text{fit}$) and $\sigma^2_\text{s,theory}$, accounting again for the uncertainty on both parameters. This metric quantifies the goodness of a linear fit between the measured and predicted values. 
\end{enumerate}

\begin{figure*}
    \centering
    \includegraphics[width=\linewidth]{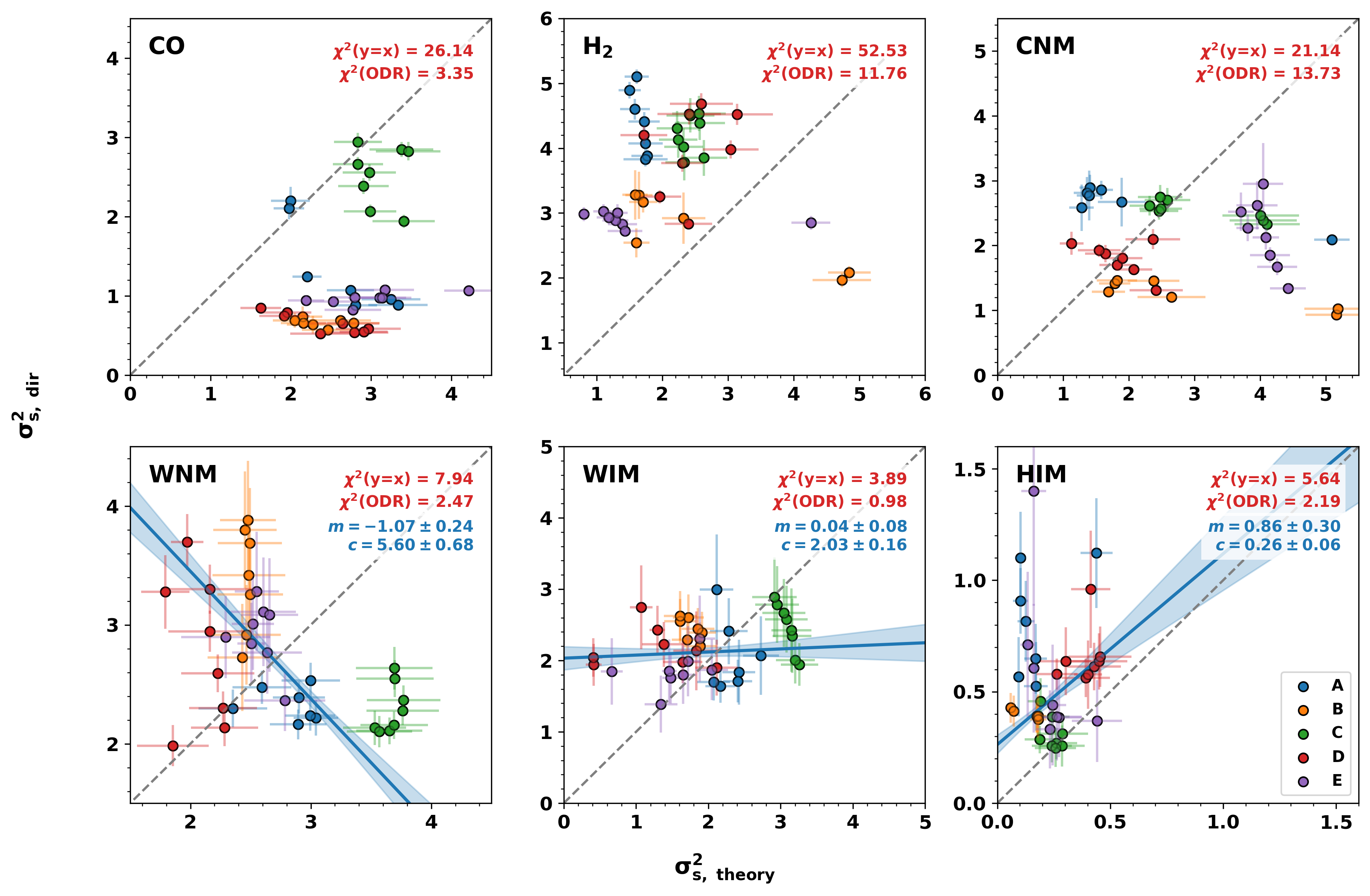}
    \caption{The width of the density PDF measured directly, $\sigma^2_{s,\mathrm{dir}}$ (see Sec.~\ref{sec:width}), versus the theoretically-expected width, $\sigma^2_{s,\mathrm{theory}}$, predicted by the isothermal $\sigma_s^2$–$\mathcal{M}$ relation, for each phase. The coloured markers indicate the values found for each molecular cloud complex at all evolutionary snapshots considered.
    The grey dashed line marks $y=x$, corresponding to perfect agreement between $\sigma^2_{s,\mathrm{dir}}$ and $\sigma^2_{s,\mathrm{theory}}$. $\chi^2_{y=x}$, listed for each phase, quantifies the deviation of the data points from this line. Also reported is $\chi^2_{\mathrm{ODR}}$, quantifying the goodness of an orthogonal distance regression (ODR) linear fit between $\sigma^2_{s,\mathrm{dir}}$ and $\sigma^2_{s,\mathrm{theory}}$  Where $\chi^2_{\mathrm{ODR}} \leq 3$, the ODR fit is shown in blue, with the surrounding shaded region indicating the $1\sigma$ confidence interval. The corresponding slope ($m$) and intercept ($c$) of the ODR fit are also reported.}
    \label{fig:theory_v_dir}
\end{figure*}

\begin{figure*}
    \centering
    \includegraphics[width=\linewidth]{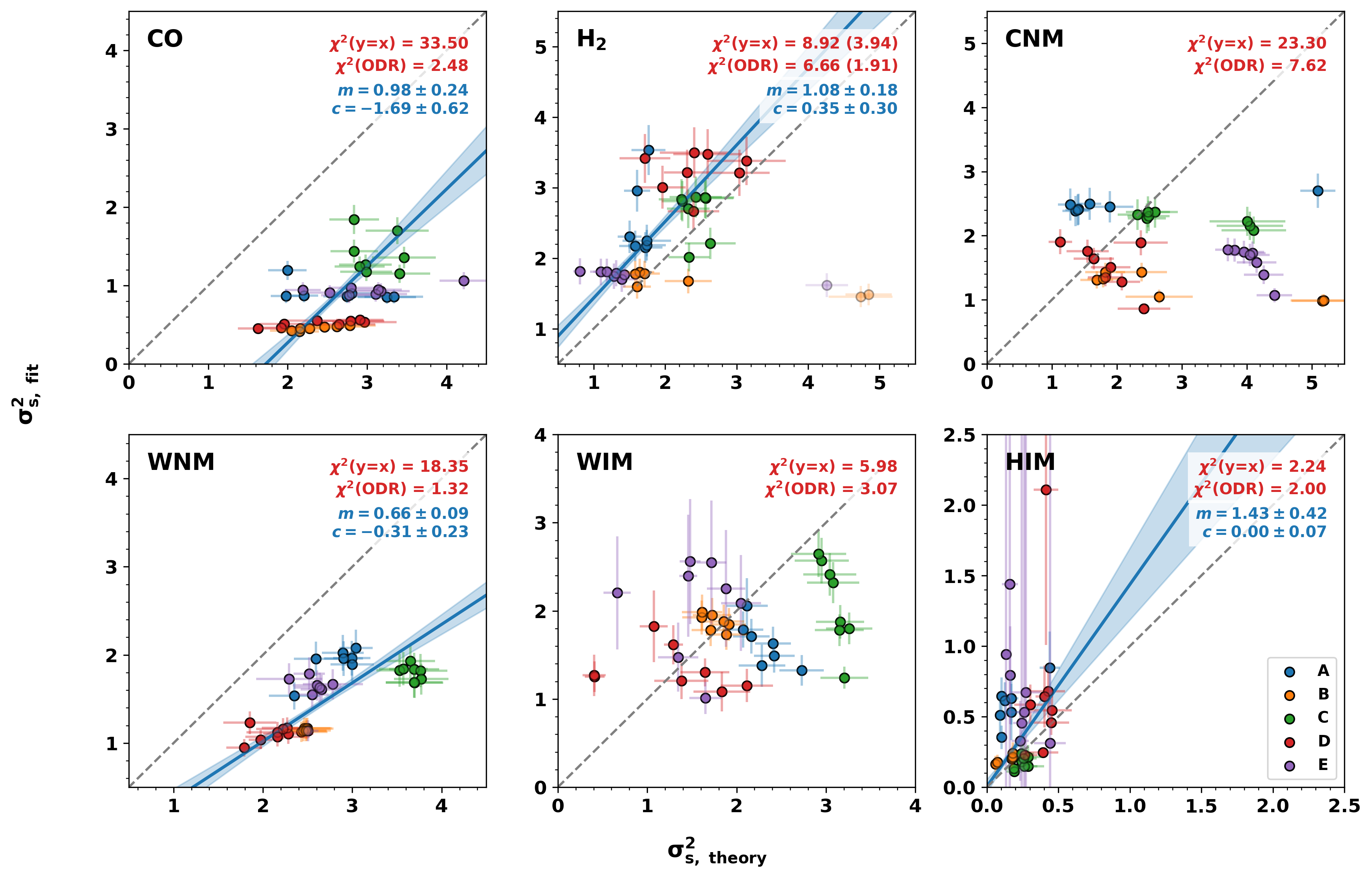}
    \caption{The width of the density PDF measured via log-normal fitting $\sigma^2_{s,\mathrm{fit}}$ (see Sec.~\ref{sec:width}), versus the theoretically-expected width, $\sigma^2_{s,\mathrm{theory}}$, predicted by the isothermal $\sigma_s^2$–$\mathcal{M}$ relation, for each phase. The coloured markers indicate the values found for each molecular cloud complex at all evolutionary snapshots considered.
    The grey dashed line marks $y=x$, corresponding to perfect agreement between $\sigma^2_{s,\mathrm{fit}}$ and $\sigma^2_{s,\mathrm{theory}}$. $\chi^2_{y=x}$, listed for each phase, quantifies the deviation of the data points from this line. Also reported is $\chi^2_{\mathrm{ODR}}$, quantifying the goodness of an orthogonal distance regression (ODR) linear fit between $\sigma^2_{s,\mathrm{fit}}$ and $\sigma^2_{s,\mathrm{theory}}$  Where $\chi^2_{\mathrm{ODR}} \leq 3$, the ODR fit is shown in blue, with the surrounding shaded region indicating the $1\sigma$ confidence interval. The corresponding slope ($m$) and intercept ($c$) of the ODR fit are also reported.}
    \label{fig:theory_v_fit}
\end{figure*}

Figures~\ref{fig:theory_v_dir} and \ref{fig:theory_v_fit} show 
$\sigma^2_\text{dir}$ versus $\sigma^2_\text{s,theory}$ and $\sigma^2_\text{fit}$ versus $\sigma^2_\text{s,theory}$ respectively, together with the corresponding values of $\chi^2_{y=x}$ and $\chi^2_\text{ODR}$. Where $\chi^2_\text{ODR}\leq3$, we consider there to be a statistically significant linear relation between the parameters and the corresponding best-fit line is shown.

\subsection{HIM and WIM}

We first examine the validity of the $\sigma^2_s-\mathcal{M}$ relation in the ionised regime. 
As noted earlier, $\sigma_\text{s,fit}\sim\sigma_\text{s,dir}$ due to the approximately log-normal shape of the density distributions of the HIM and WIM. Consequently, similar trends are observed in both Figures~\ref{fig:theory_v_dir} and \ref{fig:theory_v_fit}, with differences in $\chi^2_{y=x}$ and $\chi^2_{\text{ODR}}$ largely reflecting differences in the reported uncertainties. For the HIM, the measured values appear tentatively consistent with the values predicted by the $\sigma^2_s-\mathcal{M}$ relation. The relationship between the measured and predicted values for the WIM is more ambiguous, showing little or no clear linear trend and poor agreement between the two parameters.
In both cases, however, we are hesitant to draw conclusions as the refinement strategy of \textsc{Arepo} results in the lowest resolution in the most diffuse gas. Moreover, the lack of early-stage feedback in the Cloud Factory simulations means that the HIM and WIM are likely underestimated in the simulations and their distributions may not reflect those in the real ISM.

\subsection{WNM and CNM}

Turning our attention to the atomic gas, large discrepancies are apparent between the WNM panels in Figures~\ref{fig:theory_v_dir} and \ref{fig:theory_v_fit}. Such discrepancies arise due to the log-normal fitting procedure omitting excess low-density gas in the WNM density distributions, leading to $\sigma_\text{s,fit}^2 \lesssim \sigma_\text{s,dir}^2$. In the case of 
$\sigma^2_{s, \text{dir}}$, we observe a negative correlation between the measured and predicted values, in complete tension with the $\sigma^2_s-\mathcal{M}$ relation. 
In contrast, a positive linear relationship is present between $\sigma^2_{s, \text{fit}}$ and $\sigma^2_{s, \text{theory}}$:

\begin{equation}
    \sigma^2_{s, \text{fit}} = (0.66\pm0.09)\,\sigma^2_{s, \text{theory}}+(-0.31\pm0.23).
\end{equation}

While more consistent with theoretical expectation, the slope being less than unity indicates that the classical $\sigma^2_s-\mathcal{M}$ relation systematically overpredicts the density variance. This may indicate that additional physical processes are acting to suppress density fluctuations. One possible mechanism is non-isothermality, which alters how gas responds to compression and therefore modifies the resulting density distributions.

The impact of non-isothermality on the $\sigma^2_s-\mathcal{M}$ relation has been studied by \cite{federrath_density_2015}, who derive modified relations for polytropic equations of state, where $P \propto \rho^\Gamma$. Here, $P$ denotes the pressure, $\rho$ the density and $\Gamma$ the polytropic index, quantifying how a gas responds to compression. For a soft equation of state ($\Gamma < 1$), the gas efficiently radiates heat, allowing for stronger compression. In contrast, for a stiff equation of state ($\Gamma > 1$), inefficient cooling leads to thermal pressure that resists compression. $\Gamma =1$ corresponds to an isothermal gas. 

\cite{federrath_density_2015} derive the following modified $\sigma^2_s-\mathcal{M}$ relations for $\Gamma = 1/2$ and $\Gamma =2 $ respectively:

\begin{equation}
    \sigma^2_s = \ln \Bigg[ 1 + \frac{1}{8} \Big(
4 b^2 \mathcal{M}^2 + b^4 \mathcal{M}^4 
+ b^3 \mathcal{M}^3 \sqrt{8 + b^2 \mathcal{M}^2} \Big) \Bigg],
\end{equation}

\begin{equation}
    \sigma^2_s = \text{ln}\left[1 + \frac{1}{2}\left(-1 + \sqrt{1+8b^2\mathcal{M}^2}\right)\right].
\end{equation}

\begin{figure}
    \centering
    \includegraphics[width=0.8\linewidth]{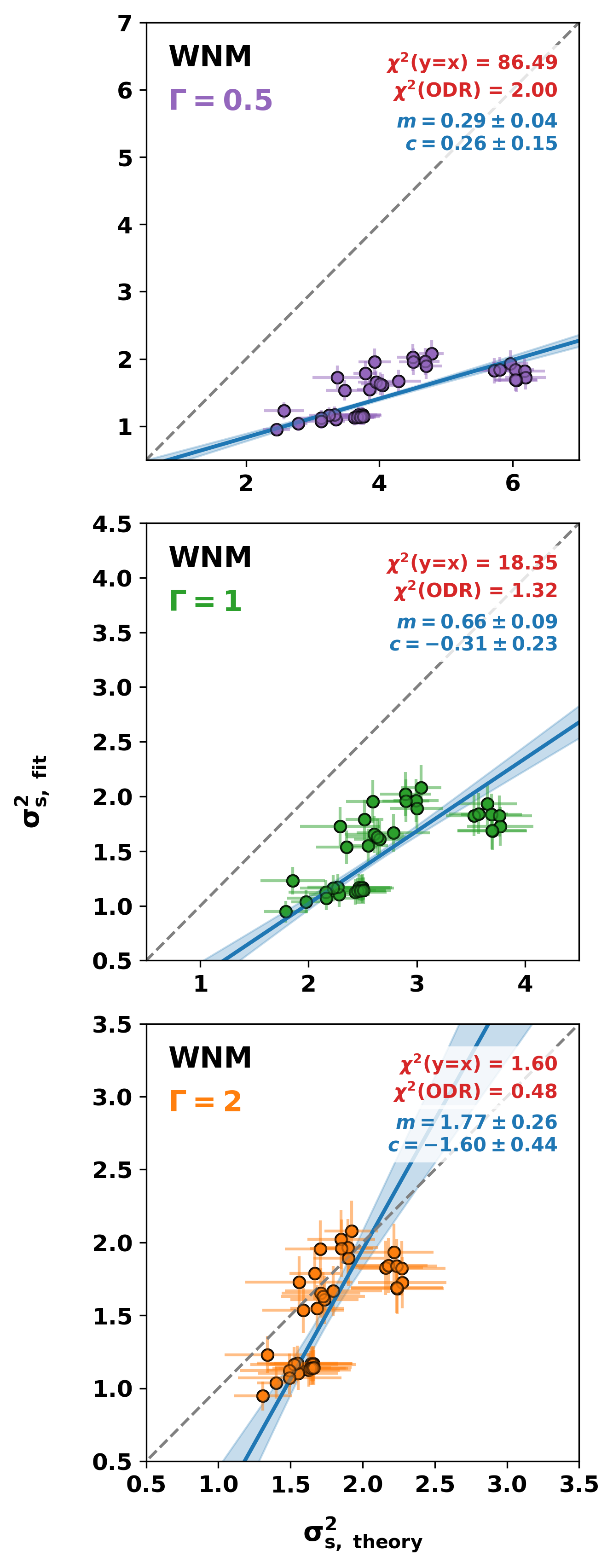}
    \caption{The width of the WNM density PDF measured via log-normal fitting, $\sigma^2_{s,\mathrm{fit}}$, versus the theoretically expected width, $\sigma^2_{s,\mathrm{theory}}$, predicted for polytropic indices $\Gamma$ = 1/2, 1 and 2. The grey dashed line marks $\sigma^2_{s,\mathrm{fit}}=\sigma^2_{s,\mathrm{theory}}$. The ODR linear fit between $\sigma^2_{s,\mathrm{fit}}$ and $\sigma^2_{s,\mathrm{theory}}$ is shown in blue, with the surrounding shaded region indicating the $1\sigma$ confidence interval.}
    \label{fig:wnm_non_isothermality}
\end{figure}

To assess whether non-isothermality can explain the widths of the WNM density PDFs in the simulations, we recompute $\sigma^2_\text{s,theory}$ using the above relations and make comparison to $\sigma^2_\text{s,fit}$ in Figure~\ref{fig:wnm_non_isothermality}. For $\Gamma=2$, the stiff equation of state smooths out density contrasts, narrowing the density PDF. This improves the agreement between the
measured and predicted values, reducing $\chi^2_{x=y}$ from 18.3 in the isothermal case to 1.6. 

\begin{figure}

    \centering
    \includegraphics[width=0.85\linewidth]{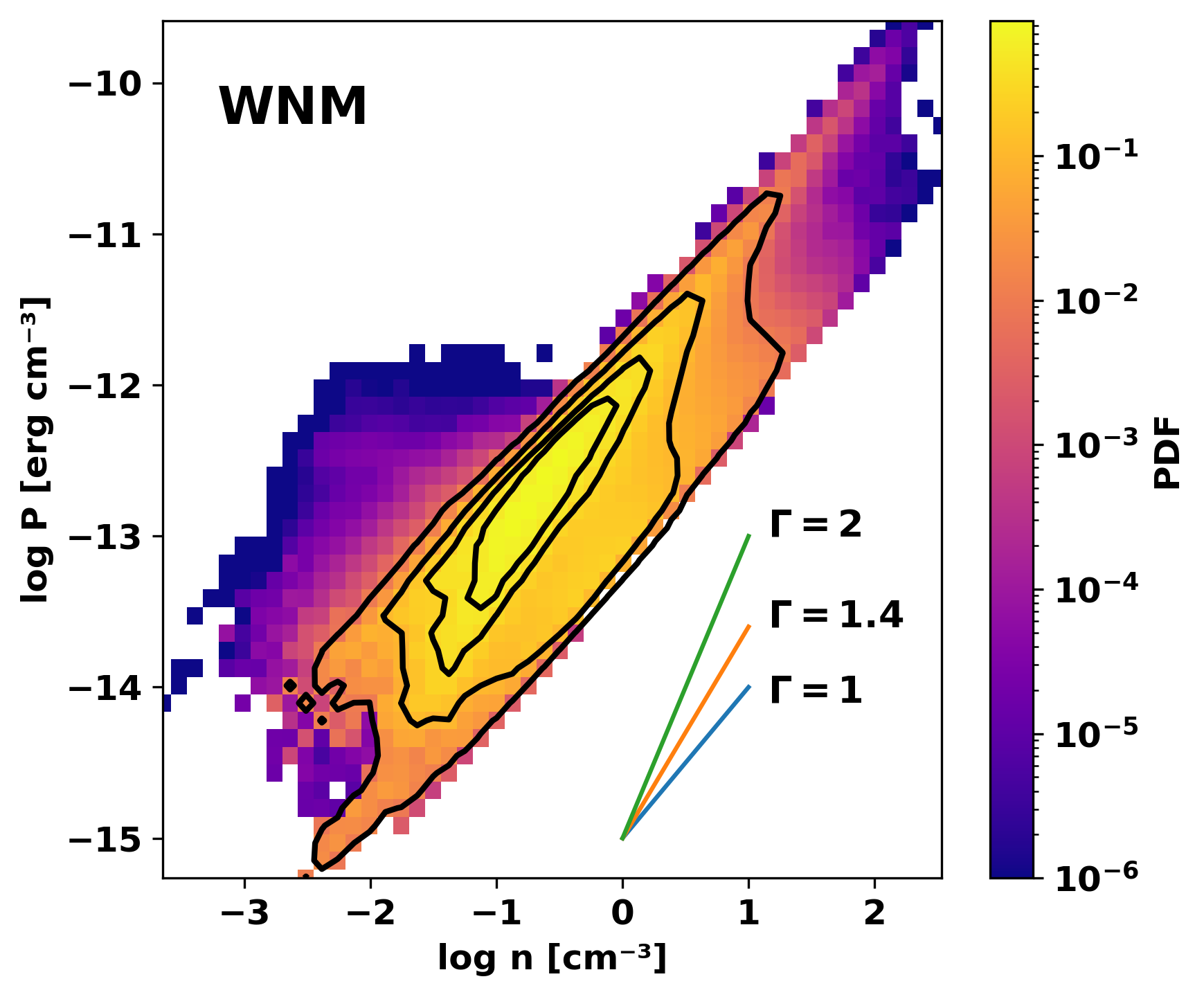}
    \caption{Pressure versus density of the WNM, for Complex C at the middle evolutionary snapshot. Each cell is weighted by the product of the cell volume and the fractional abundance of the WNM. The 2D histogram is normalised such that the bin values represent the fractional volume of the WNM. The black contours correspond to 0.01, 0.1, 0.3 and 0.5. The blue, orange and green lines show the expected pressure-density scaling for the polytropic indices $\Gamma$ = 1, 1.4 and 2 respectively.}
    \label{fig:wnm_pressure_density}
\end{figure}

A polytropic index of $\Gamma=2$ would require the WNM to cool very inefficiently. To understand whether such a stiff equation of state is a physically motivated fix, in Figure~\ref{fig:wnm_pressure_density}
we examine the pressure-density distribution of the WNM in Complex C at the middle evolutionary snapshot, where the deviation between measured and predicted values is largest. Overlaid are the pressure-density relations expected for $P \propto \rho^\Gamma$, with $\Gamma$ = 1, 1.4, and 2. While the WNM may deviate from strict isothermality, with $\Gamma > 1$ arguably providing a better description of the observed distribution, $\Gamma$ = 2 is clearly inconsistent with the data. Supporting this conclusion, \cite{federrath_density_2015} find that gas governed by such stiff equations of state exhibit highly skewed density distributions; an effect we do not observe for the WNM. Thus, while non-isothermality may partially explain the observed discrepancy with the $\sigma^2_s-\mathcal{M}$ relation, it cannot fully account for it.

Alternatively, the overprediction of the width of the WNM density PDFs may stem from an overestimation of the turbulent Mach number. Where the \textsc{hdbscan} clustering algorithm identifies few distinct structures, as is the case for the smoothly distributed WNM, $\mathcal{M}_\text{local}$ becomes sensitive to coherent motions on the scale of the simulation box, including from large-scale galactic flows.
This effect may also explain the CNM results, where neither $\sigma^2_\text{s,dir}$ nor $\sigma^2_\text{s,fit}$ show a correlation with $\sigma^2_\text{s,theory}$. Indeed, the largest outliers in the CNM panels correspond to snapshots where $\mathcal{M}_\text{local}$ approaches $\mathcal{M_\text{global}}$ (see Figure~\ref{fig:mach_global_v_local}), indicating that few structures are identified and thus $\mathcal{M}_\text{local}$ may be artificially inflated.

To test this hypothesis, we fit a linear velocity gradient to each ISM phase, modelling the bulk flow as:

\begin{equation}
    \textbf{v}_\text{bulk}(\textbf{x}) = \textbf{v}_0 + \textbf{G} \cdot (\textbf{x} - \textbf{x}_0),
\end{equation}

where $\textbf{v}_0$ and $\textbf{x}_0$ are the weighted mean velocity and position of the phase, and $\textbf{G}$ is the velocity gradient tensor. Cells are again weighted by $w_c$ (Equation~\ref{eq:weighting}). Subtracting the bulk flow from the velocity field yields the residual turbulent velocities, which are then used to recompute $\mathcal{M}_\mathrm{local}$ and the corresponding $\sigma^2_{s,\mathrm{theory}}$. 

\begin{figure}
    \centering
    \includegraphics[width=0.8\linewidth]{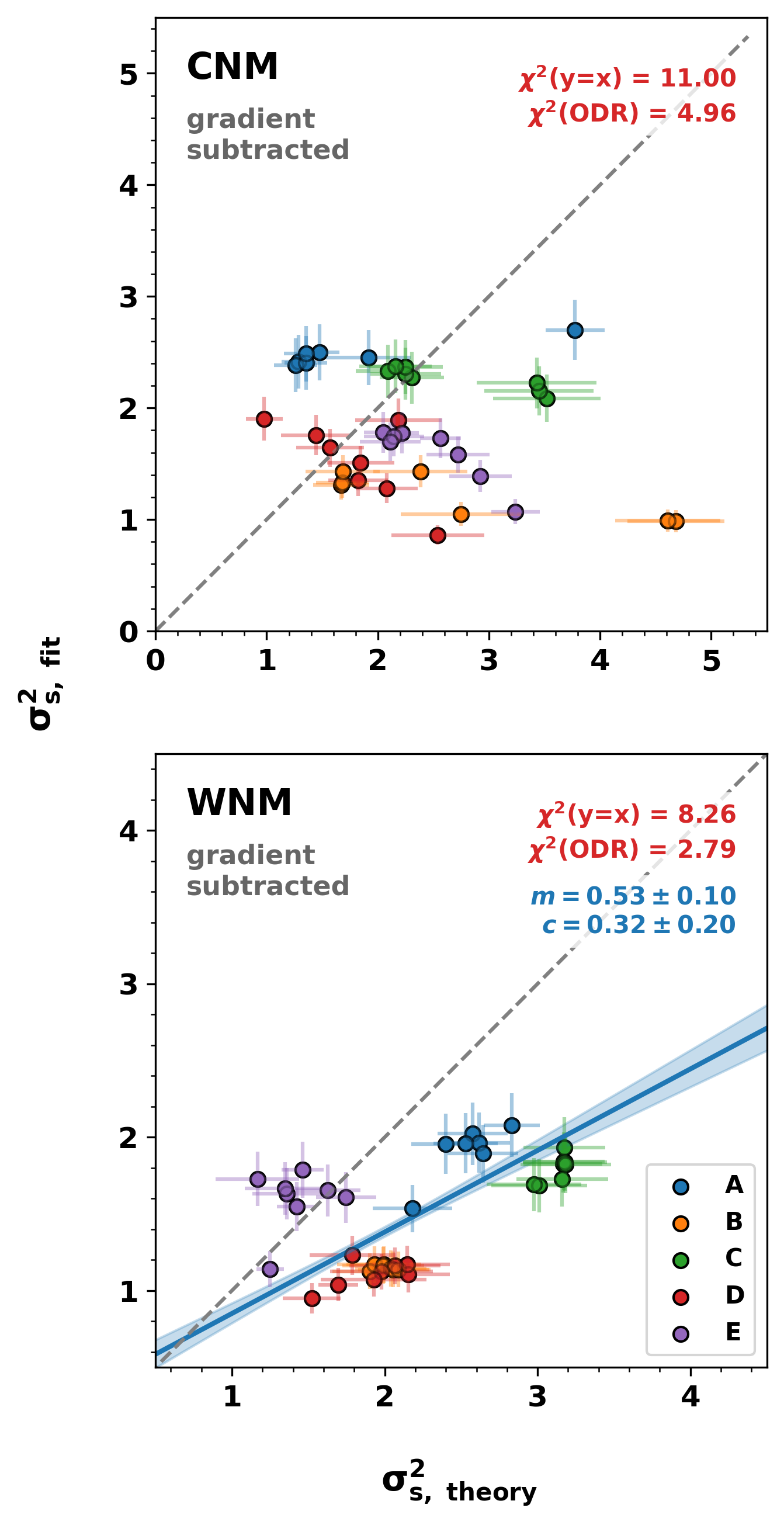}
    \caption{The width of the density PDF measured via log-normal fitting $\sigma^2_{s,\mathrm{fit}}$, versus the theoretically expected width, $\sigma^2_{s,\mathrm{theory}}$, now calculated having subtracted a linear gradient from the velocity fields.  The grey dashed line marks $\sigma^2_{s,\mathrm{fit}}$ = $\sigma^2_{s,\mathrm{theory}}$.  The ODR linear fit between $\sigma^2_{s,\mathrm{fit}}$ and $\sigma^2_{s,\mathrm{theory}}$ is shown in blue where $\chi^2_\text{ODR}\leq3$, with the surrounding shaded region indicating the $1\sigma$ confidence interval.}
    \label{fig:linear_velocity_sub}
\end{figure}

Figure~\ref{fig:linear_velocity_sub} shows $\sigma^2_\text{s,fit}$ versus the updated $\sigma^2_\text{s,theory}$ for the WNM and CNM. While subtracting a linear velocity gradient does improve the agreement between measured and predicted values, discrepancies remain. It is also worth noting that removing such large-scale gradients from the turbulent Mach number estimate may not be strictly physical. At the scale of the simulation boxes (200\,pc), such motions could plausibly be part of the turbulent cascade. Consistent with this interpretation, the power spectrum shown in Figure~\ref{fig:scale_dependence_b} shows no turnover at low $k$, indicating that the simulations probe scales below the injection scale.

In summary, the isothermal $\sigma^2_{s}-\mathcal{M}$ relation fails to capture the behaviour of both the warm and cold atomic gas. While it is perhaps unsurprising that our more sophisticated simulations are in tension with the results from isothermal periodic box set ups, identifying the physical origin of these discrepancies is non-trivial, with non-isothermality and large-scale dynamics unable to completely reconcile measured and predicted values.

\subsection{H$_2$ and CO}

Finally, we examine the $\sigma^2_s-\mathcal{M}$ relation in the molecular regime, which is of particular importance given the well-established connection between the density PDF of molecular gas and star formation. Comparing the H$_2$ panels of Figures~\ref{fig:theory_v_dir} and \ref{fig:theory_v_fit}, significant differences are observed, which arise from the omission of the gravitationally-induced power-law tail in the fitted measure of the density variance. As a result, $\sigma^2_{s,\text{fit}}$ is in considerably better agreement with $\sigma^2_{s,\text{theory}}$ than $\sigma^2_{s,\text{dir}}$, with $\chi^2_\text{ODR}$ reducing from 52 to 6.7.

Despite this improvement, three prominent outliers remain. Closer examination of these outliers reveals them to correspond to early evolutionary snapshots, suggesting that insufficient H$_2$ may have formed, leading to unreliable derived quantities (the associated Mach numbers are also outliers in  Figure~\ref{fig:mach_global_v_local}). Removing these outliers yields a statistically significant relation between $\sigma^2_{s,\text{fit}}$ and $\sigma^2_{s,\text{theory}}$:

\begin{equation}
    \sigma^2_{s, \text{fit}} = (1.08\pm0.18)\,\sigma^2_{s, \text{theory}}+(-0.35\pm0.30), 
\end{equation}

with a slope consistent with unity and a small, marginally significant offset. This suggests the classical $\sigma^2_s$–$\mathcal{M}$ relation provides a good description of the density variance of the log-normal component of H$_2$ density PDFs, provided sufficient molecular gas is present.

Lastly, we examine the results for CO. As with H$_2$, a linear relationship between the measured and predicted values emerges only in the case of the density variance of the log-normal portion of the density PDF, where

\begin{equation}
    \sigma^2_{s, \text{fit}} = (0.98\pm0.24)\,\sigma^2_{s, \text{theory}}+(-1.69\pm0.62).
\end{equation}

Although the slope is near unity, the negative offset indicates that the classical $\sigma^2_s-\mathcal{M}$ relation systematically overpredicts $\sigma^2_\text{s,fit}$. We attribute this overprediction to two effects tied to the selective nature of CO as a molecular gas tracer. First, CO is confined to the highly-shielded interiors of molecular clouds where temperatures are lowest. As a result, the average sound speed in CO-traced gas is substantially lower than across the total molecular phase ($\sim$0.3 kms\,$^{-1}$ versus $\sim$1 km\,s$^{-1}$, see Figure~\ref{fig:cs_v_vel_disp}), biasing the Mach number to higher values and increasing the predicted density PDF width. Second, photodissociation at the outer edges of molecular clouds removes CO from low-density regions, truncating the density distribution and producing a narrower density PDF than is seen for H$_2$, as illustrated in Figure~\ref{fig:density_pdfs_CO_scaled}. Together, these effects account for the breakdown of the classical $\sigma^2_s$–$\mathcal{M}$ relation, despite it holding for H$_2$, which CO is commonly accepted to trace.

\section{Conclusion}
\label{sec:conclusion}

In this study, we have used detailed zoom-in simulations of molecular clouds, modelling time-dependent chemistry, a large-scale gravitational potential, self-gravity, star formation, and supernova feedback, to examine whether the classical $\sigma^2_s-\mathcal{M}$ relation holds within individual phases of the multi-phase ISM.

Unsurprisingly, our results are strongly dependent on how key parameters are defined; an exercise that becomes increasingly ambiguous in a multi-phase and multi-scale environment. We find that adopting a `local' definition of the Mach number ($\mathcal{M}_\text{local}$) of each phase, which accounts for the spatial distribution of the gas, yields estimates more consistent with observational expectations, particularly for the highly inhomogeneous CO and H$_2$. The turbulent driving parameter, $b$, is inferred from the velocity field via Helmholtz decomposition and is found to be insensitive to scale.

Using $\mathcal{M}_\text{local}$ and $b$, we compute the density variance predicted by the isothermal $\sigma^2_s-\mathcal{M}$ relation for each phase, $\sigma^2_\text{s,theory} = \text{ln}(1 + b^2\mathcal{M}^2)$, and compare to measured values from the simulations. For each phase, the density variance is measured in two ways: the total variance of the density distribution and the variance of a fitted log-normal component. Focussing on the log-normal portion of the PDF isolates the `turbulence-driven' density variance, separating it from contributions from e.g. gravitational collapse, which are not expected to be captured by the classical relation. 

Indeed, the performance of the $\sigma^2_s-\mathcal{M}$ relation improves substantially when applied to only the log-normal component of the density PDF. Based on the comparison of measured and predicted values, we draw the following conclusions regarding the applicability of the $\sigma^2_s-\mathcal{M}$ relation in different phases of the ISM:

\begin{itemize}
    \item \textbf{WNM:} The classical $\sigma^2_s-\mathcal{M}$ relation systematically overpredicts the density variance of the log-normal component of the warm neutral medium. While the WNM exhibits some deviation from strict isothermality ($\Gamma>1$), the stiff equation of state required to fully reconcile the predicted and measured variances is inconsistent with the observed pressure-density distribution. Large-scale coherent motions may also contribute to the discrepancy, but again cannot fully capture the observed behaviour.
    \item \textbf{CNM:} The $\sigma^2_s-\mathcal{M}$ relation fails to describe the density variance of the cold neutral medium. Removing large-scale velocity gradients, as in the case of the WNM, improves the agreement between predicted and measured density PDF widths but again cannot fully resolve the tension.
    \item \textbf{H$\mathbf{_2}$:} The $\sigma^2_s-\mathcal{M}$ relation provides a good description of the width of the log-normal component of the H$_2$ density PDF. The total variance, which includes the power-law tail arising from gravitational collapse, is not captured by the relation.
    \item \textbf{CO:} The classical $\sigma^2_s-\mathcal{M}$ relation systematically overpredicts the density variance of the log-normal component of the CO density distribution due to the selective nature of CO as a tracer. With CO tracing only the coldest regions of molecular clouds, the Mach number is biased towards higher values relative to the total molecular phase, increasing the predicted density distribution width. At the same time, CO photodissociation at low densities truncates the density PDF, leading to a narrower density PDF.
\end{itemize}

We draw no conclusions for the HIM and the WIM due to poor resolution in low-density gas and the absence of early-stage stellar feedback, leaving a detailed exploration of the ionised regime to a future study. 

Ultimately, this paper has shown that the classical $\sigma^2_s-\mathcal{M}$ holds in only very specific circumstances, namely in the molecular phase where care has been taken to isolate turbulence from other processes, such as gravitational collapse and bulk motions, in the measure of the density variance and the Mach number. Outside of these conditions, the relationship breaks down. More broadly, this study has highlighted the need for caution when applying results from idealised simulations to the real ISM, and the value of detailed, high-resolution molecular cloud simulations to understand turbulence in a complex galactic environment. 

\begin{acknowledgements}

MN thanks Henrik Beuther for his supervision and for useful discussions during the writing of this paper. MN is a fellow of the International Max Planck Research School for Astronomy and Cosmic Physics at the University of Heidelberg (IMPRS-HD).
RJS gratefully acknowledges computing time provided as part of STFC DiRAC thematic project APP30360. This work used the DiRAC@Durham facility managed by the Institute for Computational Cosmology on behalf of the STFC DiRAC HPC Facility (www.dirac.ac.uk). The equipment was funded by BEIS capital funding via STFC capital grants ST/P002293/1, ST/R002371/1 and ST/S002502/1, Durham University and STFC operations grant ST/R000832/1. DiRAC is part of the National e-Infrastructure.
\end{acknowledgements}

%

\bibliographystyle{aa}
\bibliography{density_pdfs}







   
  



\begin{appendix}

\section{Impact of interpolation resolution and windowing on Helmholtz decomposition}
\label{sec:appendix_helmholtz}

\begin{figure*}[!b]
    \centering
    \includegraphics[width=0.85\linewidth]{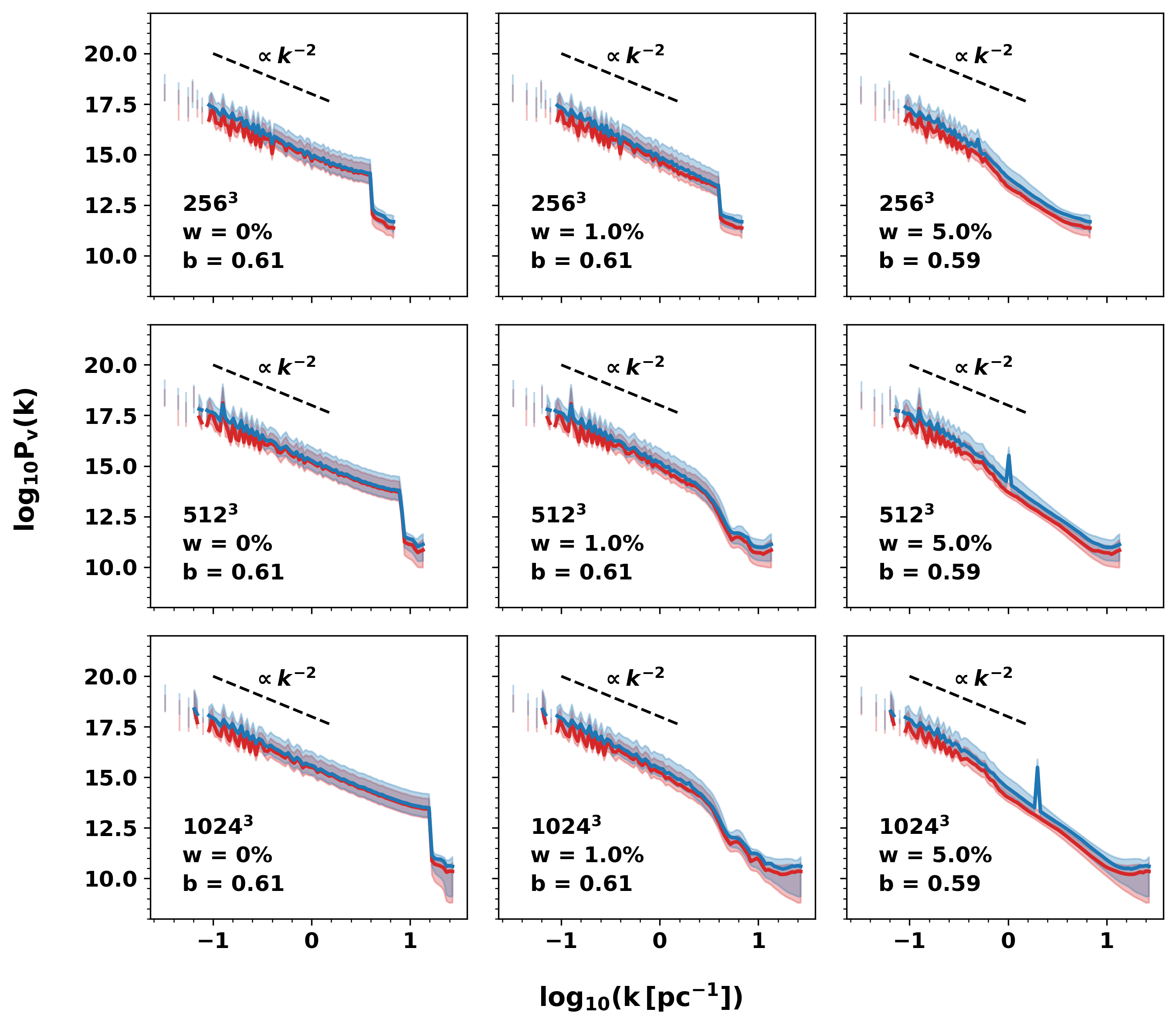}
    \caption{The 1D  time-averaged power spectra of the compressive (red line) and solenoidal (blue line) velocity components of Complex A, with corresponding shaded areas illustrating the variation over the time period studied. The spectra are plotted up to the Nyquist frequency. The black dashed line shows the expected slope of the velocity power spectrum for Burgers turbulence. The power spectra are plotted for varying resolutions of Cartesian grids, onto which the native \textsc{Arepo} data is interpolated, and different Hanning window sizes, with $w$ the percentage of the box the window is applied to. $b$ is the time-averaged, turbulent driving parameter for each setup.}
    \label{fig:res_window_testing}
\end{figure*}


Here we examine the impact of the interpolation resolution and windowing on the turbulent driving parameter, $b$, inferred from the velocity fields via Helmholtz decomposition. As well as the original $512^3$ grid used for the main results of this paper, we also interpolated the native \textsc{Arepo} velocity fields to Cartesian grids of resolution $256^3$ ($\Delta=0.78$\,pc) and $1024^3$ ($\Delta=0.20$\,pc). We applied Hanning windows to the grids to account for the non-periodicity of the simulation boxes. 
For each resolution, we tested three cases: no window, a window spanning 1$\%$ of the grid length, and a window spanning 5$\%$ of the grid length.
For each setup, we calculated $b$, as well as the 1D time-averaged power spectra of the compressive and solenoidal velocity components. The results are shown in Figure~\ref{fig:res_window_testing} for Complex A, with the power spectra plotted up to the Nyquist frequency for the given resolution. 

\newpage

We find that $b$ is largely insensitive to the interpolation resolution and windowing, with the time-averaged value changing only slightly from 0.61 to 0.59 in the case of the largest window. In contrast, the behaviour of the velocity power spectra at small scales is dependent on both the resolution of the Cartesian grid and the size of the applied Hanning window. The downturn at high $k$ is observed to shift to smaller $k$-modes as the resolution increases, reflecting the suppression of high $k$-modes when small-scale structure is no longer captured by a coarse interpolation grid. Applying a window further suppresses power at small scales through smoothing the velocity field at the grid boundaries. This demonstrates that the location of the downturn is not associated with a dissipation scale in the simulations, but rather arises from numerical choices made during the interpolation and Fourier analysis. As such, the downturn should not be interpreted as the scale at which turbulence is no longer resolved in the simulations.






\FloatBarrier 
\clearpage

\end{appendix}
\end{document}